\documentclass[twocolumn,secnumarabic, 
, amssymb, aps, prl, superscriptaddress]{revtex4-1}

\usepackage{graphicx}
\usepackage{amsmath}
\begin{document}

\title{Electric field control of radiative heat transfer in a superconducting circuit}%

\author{Olivier Maillet}%
\email{olivier.maillet@aalto.fi}
\affiliation{QTF Centre of Excellence, Department of Applied Physics, Aalto University School of Science, P.O. Box 13500, 00076 Aalto, Finland}
\author{Diego A. Subero Rengel}%
\affiliation{QTF Centre of Excellence, Department of Applied Physics, Aalto University School of Science, P.O. Box 13500, 00076 Aalto, Finland}
\author{Joonas T. Peltonen}%
\affiliation{QTF Centre of Excellence, Department of Applied Physics, Aalto University School of Science, P.O. Box 13500, 00076 Aalto, Finland}
\author{Dmitry S. Golubev}%
\affiliation{QTF Centre of Excellence, Department of Applied Physics, Aalto University School of Science, P.O. Box 13500, 00076 Aalto, Finland}
\author{Jukka P. Pekola}%
\affiliation{QTF Centre of Excellence, Department of Applied Physics, Aalto University School of Science, P.O. Box 13500, 00076 Aalto, Finland}
\date{June 8, 2020}%
\maketitle

\textbf{
	Heat is detrimental for the operation of quantum systems, yet it fundamentally behaves according to quantum mechanics, being phase coherent and universally quantum-limited regardless of its carriers. Due to their robustness, superconducting circuits integrating dissipative elements are ideal candidates to emulate many-body phenomena in quantum heat transport, hitherto scarcely explored experimentally. However, their ability to tackle the underlying full physical richness is severely hindered by the exclusive use of a magnetic flux as a control parameter and requires complementary approaches.
	Here, we introduce a dual, magnetic field-free circuit where charge quantization in a superconducting island enables thorough electric field control. We thus tune the thermal conductance, close to its quantum limit, of a single photonic channel between two mesoscopic reservoirs. We observe heat flow oscillations originating from the competition between Cooper-pair tunnelling and Coulomb repulsion in the island, well captured by a simple model. Our results highlight the consequences of charge-phase conjugation on heat transport, with promising applications in thermal management of quantum devices and design of microbolometers.
}
\section{Introduction}
\begin{figure*}
	\includegraphics[width=17.2cm]{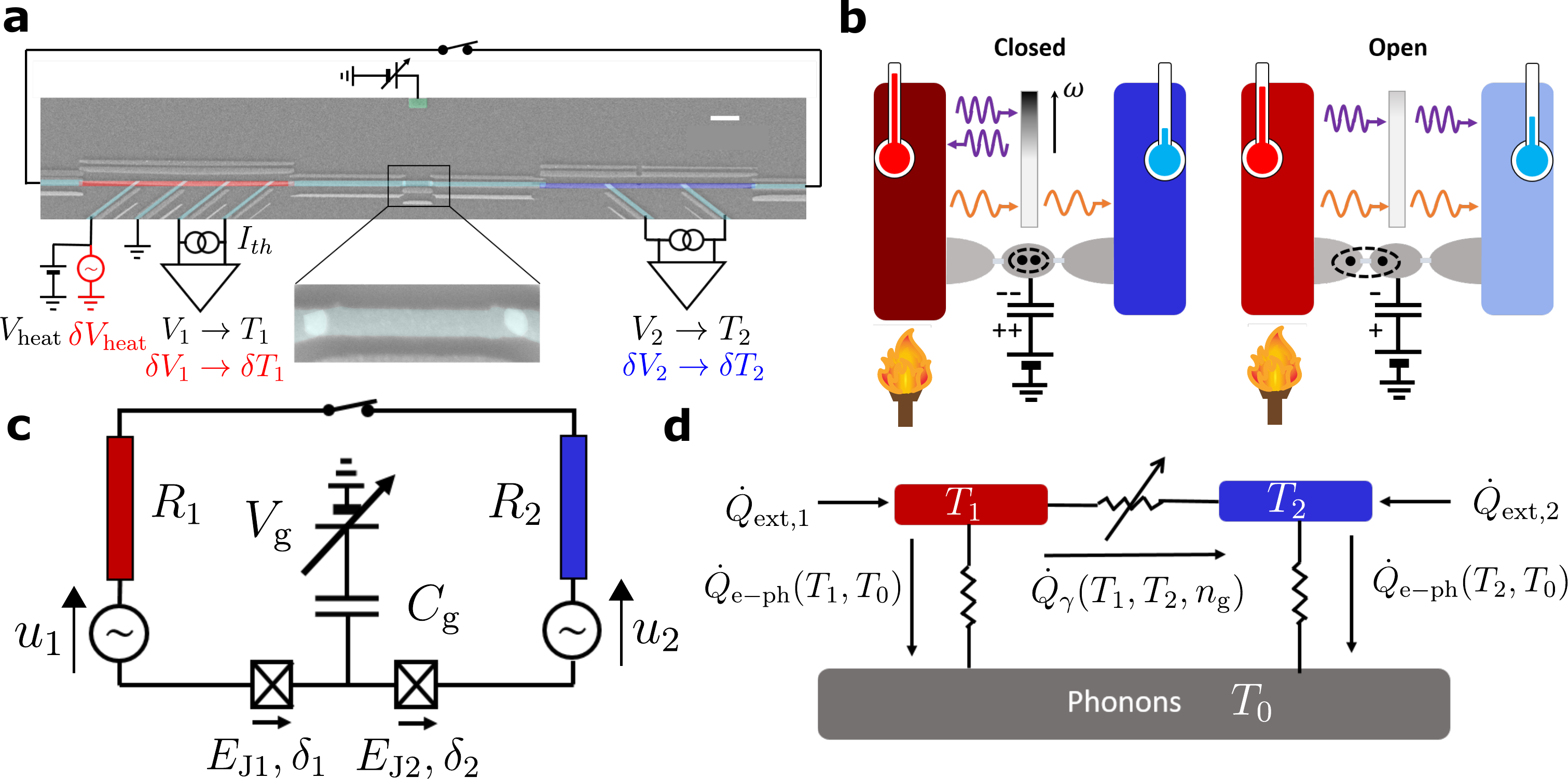}
	\caption{\textbf{Experimental setup and principle of the electrostatic photonic heat valve.} \textbf{a} Scanning electron micrograph (scale bar: 2 $\mu$m) of the Cooper pair transistor (central part with close-up), controlled with a side gate electrode (upper green lead) and connected via clean contacts and superconducting aluminium strips (light blue) to copper thin films (red and blue). Aluminium leads (oblique, light blue) are connected via oxide tunnel barriers to the films for local Joule heating (AC $\delta V_{\mathrm{heat}}$ and DC $V_{\mathrm{heat}}$) and/or electronic thermometry (using DC floating current sources and voltage amplifiers (resp. lock-in amplifiers) for DC ($V_{1,2}$) [resp. AC ($\delta V_{1,2}$)] read-out). A millimetric aluminium bonding wire closes the circuit, and can be removed to perform characterization and control measurements, hence the switch representation. \textbf{b} Principle of the experiment: the resistors are kept under constant temperature gradient, and emit Johnson noise. The CPT acts as a filter for radiated photons, with a frequency-dependent transmission coefficient schematically represented as a color gradient where the darker, the less transparent. In closed position, a single Cooper pair is localized in the superconducting island by applying an induced charge of $2e$ through the gate voltage, which reduces the noise current and hence the heat transfer. In open position, the $1e$ induced charge favours Cooper-pair tunneling and therefore increases the bandwidth of noise current, leading to increased heat transfer. \textbf{c} Electrical representation of the central circuit, with Johnson-Nyquist voltage noises $u_{1,2}$ represented as sources in series with the resistors $R_{1,2}$, $\delta_{1,2}$ the phases accross the two Josephson junctions and $E_{\mathrm{J}1,2}$ their Josephson energies. \textbf{d} Schematic representation of the thermal balance of the system.}
	\label{fig1}
\end{figure*}

Ohmic resistors embedded in mesoscopic superconducting circuits are well suited to the study of radiative transfer physics, due to the correspondence between Planck's black-body radiation and Johnson-Nyquist noise originating from a resistive element \cite{Johnson1928,Nyquist1928}. Consider an arbitrary electrical circuit connecting two resistors $R_1,R_2$ kept at different temperatures $T_1,T_2$. Their voltage noises, which simply arise from black-body photon emission/absorption events, result in a global noise current flowing in the circuit, leading to Joule dissipation by the resistive elements. In the lumped element approximation, valid at low temperatures in the case of a millimeter scale circuit such as the one depicted in Fig. \ref{fig1}a), the thermal photon wavelength $hc/k_{\mathrm{B}}T\approx 10~\mathrm{cm}$ at $150~\mathrm{mK}$ is bigger than the typical size of the circuit. Thus the power transmission coefficient $\tau$ between the two resistances can be made explicit using a standard circuit approach \cite{Schmidt2004,Pascal2011}: $\tau(\omega)=4R_1R_2/|Z_{\mathrm{tot}}(\omega)|^2$, where $Z_{\mathrm{tot}}(\omega)$ is the total circuit series impedance at angular frequency $\omega$. The net power $\dot{Q}_{\gamma}$ radiated from the hot to the cold resistor writes
\begin{equation}
\label{heat_Landauer}
\dot{Q}_{\gamma}=\int_{0}^{\infty}\frac{\mathrm{d}\omega}{2\pi}\tau(\omega)\hbar\omega\left[n_1(\omega)-n_2(\omega)\right].
\end{equation}
Here $n_{i}(\omega)=1/\left[\exp(\hbar\omega/k_{\mathrm{B}}T_i)-1\right]$ is the thermal population of the reservoir $i$, i.e. its Bose distribution at temperature $T_i$. The populations determine the thermal cut-off frequency $k_{\mathrm{B}}T/\hbar$ of the radiation spectrum, which lies in the microwave range at cryogenic temperatures ($\sim 3 $ GHz at 150 mK). The maximum unity transmission leads to heat transfer at the universal quantum limit of thermal conductance $G_{\mathrm{Q}}=\pi k_{\mathrm{B}}^2T/6\hbar\approx (1~\mathrm{pW/K}^2)T$ \cite{Pendry_1983,Schwab2000,Meschke2006,Chiatti2006,Timofeev2009,Jezouin2013,Partanen2016,Banerjee2017,Cui1192}. In our electrical approach (which may be generalized to arbitrary carriers statistics within the Landauer formalism \cite{Blencowe1999}) this limit corresponds to perfect impedance matching, i.e. $R_1=R_2$ with no additional contributions over the full black-body spectral range. Adding an appropriate tunable series reactance (a heat valve) permits tuning of the transmission coefficient without adding dissipation. Up to now, theoretical proposals \cite{Ojanen2008} and realizations \cite{Meschke2006,Ronzani2018,Partanen2018} of a photonic heat valve only considered magnetic control, which is usually rather unpractical to implement. Besides, a larger degree of control may be required for fundamental investigations of heat transport in the quantum regime \cite{Giazotto2012,Saito2013,Pepa2014,Gely2019}, e.g. by manipulating simultaneously charge and flux degrees of freedom. By contrast, electric control is now well established in electronic heat transport and thermoelectricity experiments, whether using a single-electron transistor \cite{Dutta2017}, a quantum dot \cite{Scheibner2005,Thierschmann2015,Josefsson2018,Dutta2019,Jaliel2019,Dutta2020} or a quantum point contact \cite{Molenkamp1992,Chiatti2006,Jezouin2013,Sivre2017,Sivre2019}, but it has not been considered for radiative heat transport. In this Article we experimentally demonstrate a fully electrostatic photonic heat valve operating close the quantum limit: in between our two thermal baths connected by superconducting lines, we include a Cooper pair transistor (CPT) \cite{joyez1995}, a small superconducting island where electrostatic fields impede the charge transfer, a phenomenon commonly referred to as Coulomb blockade. Its magnitude can be simply adjusted by controlling the offset charge of the island with an electrostatic field via a nearby gate electrode. By varying the gate charge by an amount $e$, the effective series impedance is tuned from, ideally, matched case to mismatch, which in turn opens or closes the heat valve at will, as schematically displayed in Fig. \ref{fig1}b).

\section{Results}
\textit{Experimental setup and principle.} The sample [see Fig. \ref{fig1}a) for a micrograph and c) for the equivalent circuit] is measured in a dilution refrigerator, and addressed with filtered lines to minimize external noise. The system is a series combination of two nominally identical, small ($100\times 100~\mathrm{nm}^2$) Josephson junctions delimiting a small island of dimensions $1.4~\mu\mathrm{m}\times 170~\mathrm{nm}\times 22~\mathrm{nm}$ with capacitance $C$, forming the Cooper pair transistor. This ensemble is attached on both sides to nominally identical thin copper films of volume $\Omega=10~\mu\mathrm{m}\times 200~\mathrm{nm}\times 12~\mathrm{nm}$ chosen so as to maximize their resistance (and thus the transmission factor $\tau$) while having minimal stray capacitance and thermal gradients. These resistors thus act as quasiparticle filters for the CPT \cite{joyez1995} as well as thermal baths, referred to in the following as source and drain. The clean electrical contact between the superconducting (S) circuit line and the normal metal (N) resistor acts as an Andreev mirror \cite{Andreev1964}, transmitting charge but preventing heat carried by quasiparticles from flowing outside the resistors. As a result, electronic heat transport by quasiparticles is efficiently suppressed along the superconducting line at dilution temperatures \cite{Timofeev2009}. The ensemble is electrically closed into a loop by a short ($\sim 5$ mm) Al bonding wire to ensure that noise current carrying the photonic heat does flow and remain integrally in the so formed floating circuit \cite{Timofeev2009}. Superconducting leads are connected to the resistors through thin oxide barriers. These Normal-Insulator-Superconductor (NIS) tunnel junctions enable one to measure the quasi-equilibrium electronic temperature of the resistor or to locally tune it via Joule heating \cite{Giazotto2006}. Transport measurements made in a run prior to closing the loop (see Supplementary Methods) yield the resistances values $R_i\approx 290\pm 30~\Omega$, the gate capacitance $C_{\mathrm{g}}\approx 12$ aF, CPT (single electron) charging energy $E_{\mathrm{c}}=e^2/2C=k_{\mathrm{B}}\times 0.64~\mathrm{K}$ and Josephson energy per junction $E_{{\mathrm{J}}}=k_{\mathrm{B}}\times 0.69~\mathrm{K}\sim E_{\mathrm{c}}$. 
\begin{figure}
	\includegraphics[width=8.6cm]{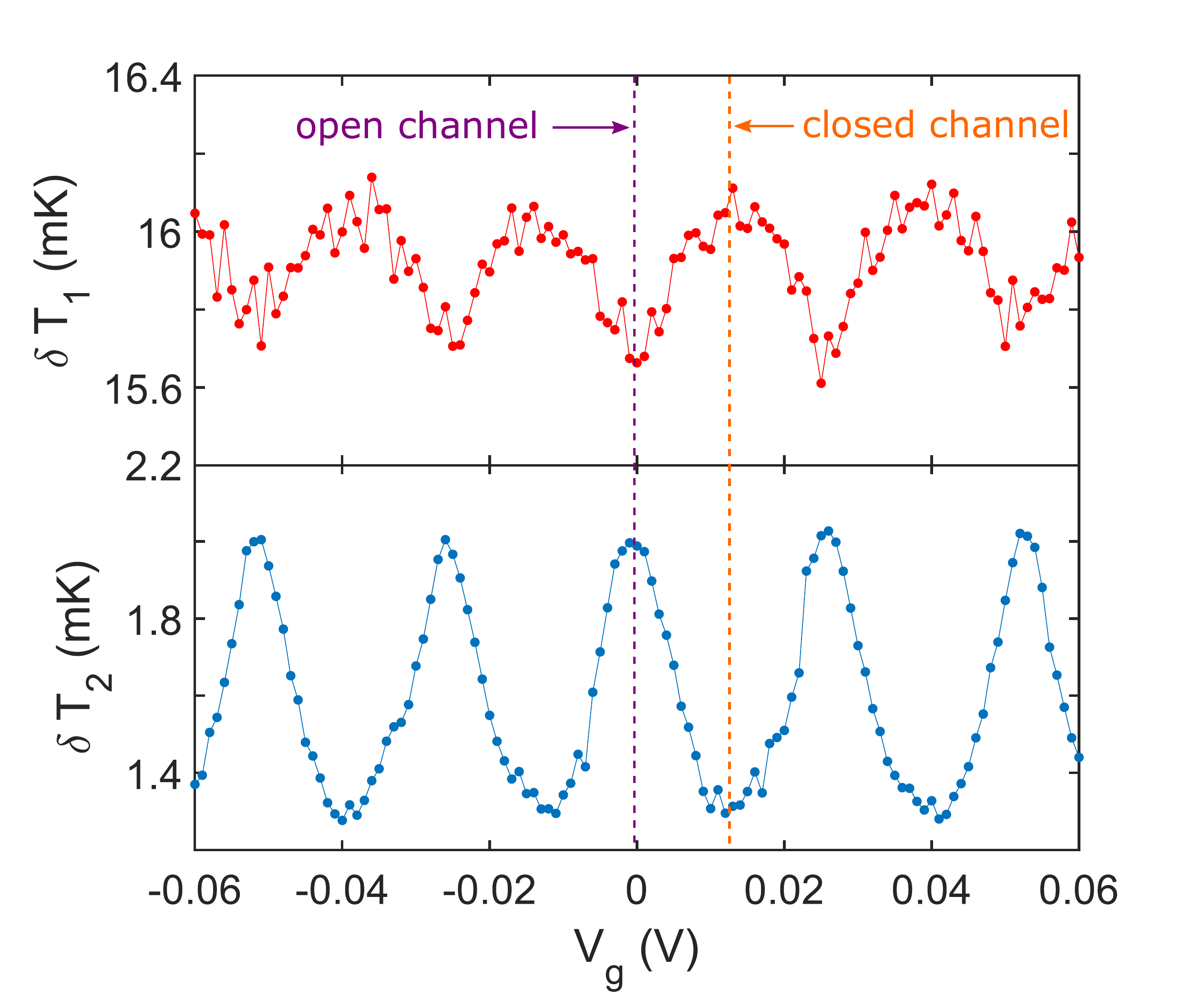}
	\caption{\textbf{AC heating measurements.} Oscillation amplitude of the temperature response to small AC heating, recorded simultaneously in the source (upper panel) at DC temperature $T_1=203$ mK and drain (lower panel) at  $T_2=170$ mK as a function of the applied gate voltage $V_{\mathrm{g}}$.}
	\label{fig2}
\end{figure}

We then investigate heat transport under an imposed temperature difference between the source and the drain. Any heat load brought externally to one of the resistors, say, the source, heats it up quickly ($\sim$ 1 ns) via electron-electron scattering to a quasi-equilibrium electronic temperature \cite{Pothier1997}, whose steady-state value is determined by taking into account two inelastic scattering mechanisms for heated electrons. The first one is electron-phonon relaxation \cite{Pinsolle2016}, which is minimized due to operation at low temperature and the small volume of the copper film. The second relaxation process occurs via electron-photon coupling \cite{Schmidt2004} and is expected to be dominant at low temperature since electron-phonon thermal conductance vanishes as $G_{\mathrm{e-ph}}\propto \Omega T^4$ \cite{Giazotto2006}. No intentional heat load is brought to the drain, and hence we can ascribe any temperature change observed there to a reservoir-reservoir heat flow, through the photonic channel. 

A diagram summarizing the different heat flows in the system is depicted in Fig. \ref{fig1}d). The thermal balance of the system in steady state for a cryostat temperature $T_0$ writes for each resistor $i$:
\begin{equation}
\label{thermal_balance}
\dot{Q}_{\mathrm{ext},i}=\dot{Q}_{\mathrm{e-ph},i}(T_i,T_0)-(-1)^{i}\dot{Q}_{\gamma}(T_1,T_2,n_{\mathrm{g}})\\
\end{equation}
where the electron-phonon heat flow for resistor $i$ is $\dot{Q}_{\mathrm{e-ph},i}=\Sigma\Omega(T_i^5-T_0^5)$, with  $\Sigma\approx 3.7\times 10^9~\mathrm{W.m}^{-3}.\mathrm{K}^{-5}$ the electron-phonon coupling constant, measured independently (see Supplementary Methods), for copper, and $\dot{Q}_{\gamma}$ is the source-drain heat flow. Using a lock-in amplifier, we measure small variations of temperature of peak amplitude $\delta T_{1,2}$ in both reservoirs upon a small sinusoidal heating at frequency $f=77$ Hz added to the DC power brought through one source NIS junction. Assuming that steady-state is valid at each modulation increment ($f$ is much smaller than any relaxation rate) and $T_1-T_2,\delta T_{1,2}\ll T_{1,2}$, from Eq. (\ref{thermal_balance}) we obtain an experimental value of thermal conductance between reservoirs 1 and 2 (see Methods):
\begin{equation}
\label{phot_heat_cond}
G_{\gamma}=\frac{5\Sigma\Omega T_2^{4}}{\delta T_1/\delta T_2-1},
\end{equation}
with $T_2$ monitored within $\pm 1~\mathrm{mK}$ with a DC voltmeter. Such an AC technique allows us to measure heat currents with a resolution down to $100~\mathrm{aW}.\mathrm{Hz}^{-1/2}$, without suffering from excessive charge noise.

\textit{Conductance modulation and model.} The temperature modulation amplitude in source and drain as a function of the applied gate voltage $V_{\mathrm{g}}$ is shown in Fig. \ref{fig2} for DC temperatures $T_1=203$ mK and $T_2=170$ mK. Clear oscillations are observed, that are $2e$-periodic in the gate charge $en_{\mathrm{g}}=C_{\mathrm{g}}V_{\mathrm{g}}$. This is a strong indication that an interplay between Cooper-pair tunneling and Coulomb blockade in the superconducting island is behind the mechanism for heat modulation, similar to the critical current modulation of the CPT \cite{joyez1995,Aumentado2004}. In addition, the temperature modulation is minimum (resp. maximum) in the source when that of the drain is maximum (resp. minimum), which can be correlated with opening (resp. closing) the photonic channel with the gate voltage. The amplitude of temperature oscillations is determined by the amount of power brought by the AC Joule heating signal, as well as the thermal balance: at our operation temperatures, electron-phonon relaxation in the reservoir is comparable with or dominant over the electron-photon mechanism. Therefore, the applied AC signal must be large enough to observe a sizeable response with good signal to noise ratio both in source and drain and to observe the gate modulation. On the other hand, it must be small enough to remain within the linear regime and keep the experimental definition of the photonic thermal conductance (\ref{phot_heat_cond}) valid. The data are taken with AC excitations kept in a range that satisfy both requirements.

\begin{figure*}
	\includegraphics[width=17.2cm]{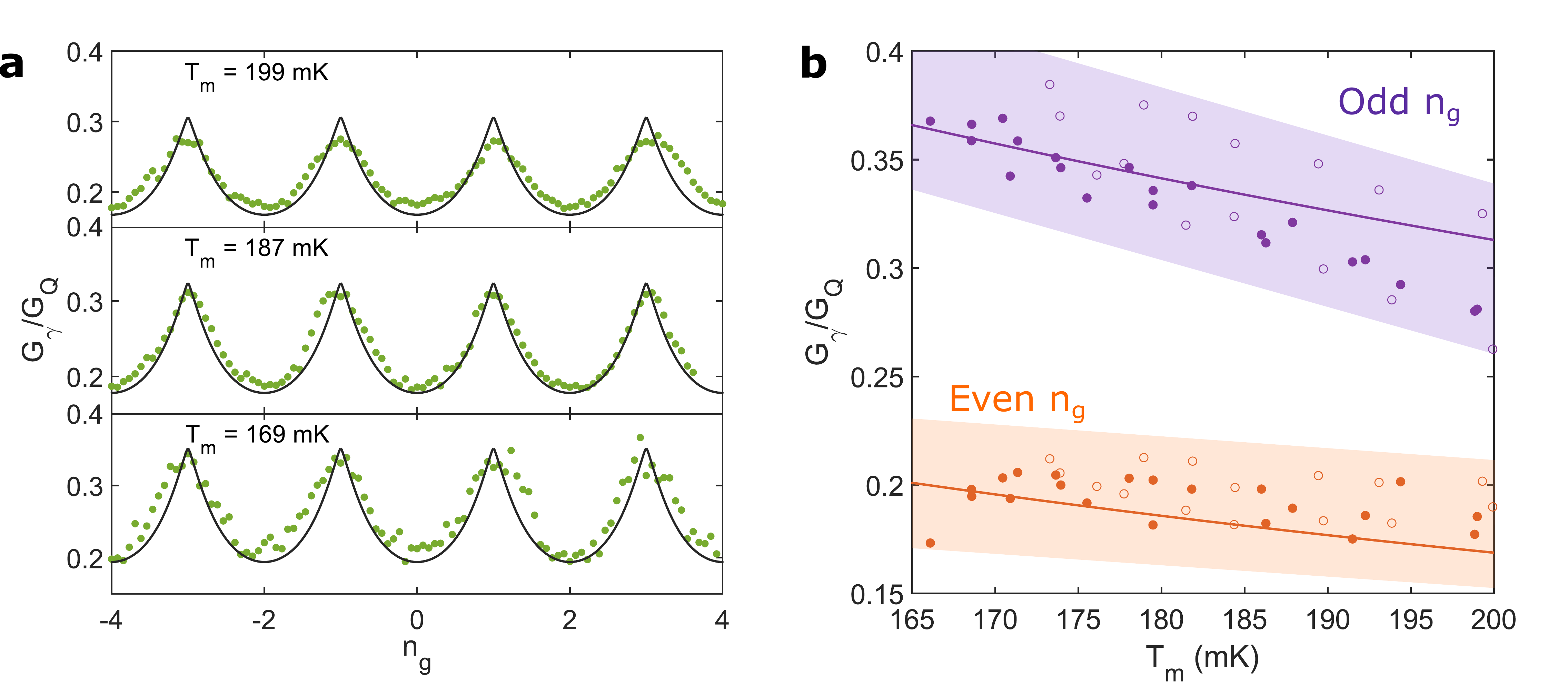}	
	\caption{\textbf{Coulomb oscillations of photonic thermal conductance and even-odd discrepancy.} \textbf{a} Source-drain thermal conductance, normalized to the quantum limit $G_{\mathrm{Q}}=\pi k_{\mathrm{B}}^2T_{\mathrm{m}}/6\hbar$, as a function of the gate charge $n_{\mathrm{g}}=C_{\mathrm{g}}V_{\mathrm{g}}/e$, for three median temperatures $(T_{\mathrm{m}}=169,187$ and $199~\mathrm{mK})$. The theoretical lines are the application of Eq. (\ref{Conductance_theo_new}). \textbf{b} Normalized conductances measured at gate open (purple dots) and closed (orange dots) positions, at cryostat temperatures 100 mK (full dots) and 150 mK (empty dots). The solid lines are the application of Eq. \eqref{Conductance_theo_new} for $n_{\mathrm{g}}=0~\mathrm{mod~2}$ (orange) and $n_{\mathrm{g}}=1~\mathrm{mod~2}$ (purple). The coloured zones delimit the typical uncertainty, based on the measurement noise and parameters uncertainties.}
	\label{fig3}
\end{figure*}

Oscillations of the source-drain thermal conductance for three mean temperatures are represented in Fig. \ref{fig3}a), normalized to the thermal conductance quantum $G_{\mathrm{Q}}$. The typical values are smaller than the conductance quantum $G_{\mathrm{Q}}$ ($\sim 35$ \% of the quantum limit at maximum), while the maximum achieved contrast $\mathcal{C}=$ $(G_{\gamma,\mathrm{max}}-G_{\gamma,\mathrm{min}})/(G_{\gamma,\mathrm{max}}+G_{\gamma,\mathrm{min}})=0.28$ is far from reaching 1, as expected from the impedance mismatch introduced by the Josephson device. The thermal conductance oscillations can be understood in terms of $2e$ quantization of the charge on the island, jointly with the charge-phase conjugation at work in the CPT \cite{joyez1995,Devoret1997}. Assuming for simplicity that the two junctions are identical and neglecting quasiparticle excitations \cite{joyez1995}, the Hamiltonian $\hat{\mathcal{H}}$ of the system writes:
\begin{equation}
\label{CPT_Hamiltonian}
\hat{\mathcal{H}}=E_{\mathrm{c}}(\hat{n}-n_{\mathrm{g}})^2-2E_{\mathrm{J}}\cos{\hat{\phi}}\cos{\frac{\delta}{2}},
\end{equation}
where $\hat{n}$ is the number operator of excess paired electrons in the island, $\hat{\phi}=\delta_2-\delta_1$ the phase of the island, and $\delta=\delta_1+\delta_2$ the total phase across the CPT. $\hat{n}$ and $\hat{\phi}$ are canonically conjugated variables whose uncertainties $\Delta n$ and $\Delta\phi$ satisfy the relation $\Delta n\Delta\phi\geq 1$ \cite{Devoret1997}. For odd values of $n_{\mathrm{g}}$ the Coulomb gap $\Delta E=4E_{\mathrm{c}}(1-n_{\mathrm{g}}\mod 2)$, which represents the electrostatic energy cost of adding one Cooper pair on the island, is closed. This leads to maximum quantum fluctuations of the charge degree of freedom $\hat{n}$ since Josephson coupling fixes the phase across the junctions and thus $\hat{\phi}$. Therefore, the Josephson supercurrent flowing across the CPT is maximum, which in turn minimizes the Josephson inductance $L_{\mathrm{J}}=\hbar/2eI_{\mathrm{C}}$, where $I_{\mathrm{C}}$ is the CPT critical current. As a result, the bandwidth for thermal excitations is increased and so is the thermal conductance. Conversely, for even values of the gate charge, the Coulomb gap is maximized, which tends to freeze the number of Cooper pairs on the island. As a result, quantum phase fluctuations are increased, thus leading to a reduction of the effective Josephson coupling and thus of the cut-off frequency for thermal currents. 

The above picture can be captured by a simple circuit model. The Josephson element under zero bias, small junction capacitance and in a low impedance environment ($R_{1,2}\ll R_{\mathrm{Q}}=h/4e^2=6.45~\mathrm{k}\Omega$) may be approximated as a gate-tunable inductor $L_{\mathrm{J}}$ in series with the resistors in the frequency range relevant for thermal photons at 100-200 mK \cite{SuppMat}. The circuit with series impedance $Z_{\mathrm{tot}}(\omega)=R_1+R_2+\mathrm{i}L_{\mathrm{J}}\omega$ is thus a low-pass filter for thermal radiation with a gate-tunable cut-off frequency $\omega_{\mathrm{c}}(n_{\mathrm{g}})=(R_1+R_2)/L_{\mathrm{J}}(n_{\mathrm{g}})$. The gate dependence of the critical current $I_{\mathrm{C}}$ is derived by finding the maximum tilt, allowed in the supercurrent branch, of the effective Josephson washboard potential modified by Coulomb interactions on the island, $I_{\mathrm{C}}(n_{\mathrm{g}})=2e/\hbar\max_{\delta}\partial E_0(n_{\mathrm{g}},\delta)/\partial\delta$,
where $E_0(n_{\mathrm{g}},\delta)$ is the ground eigenenergy band of the system (see Supplementary Note 1) obtained from (\ref{CPT_Hamiltonian}). We can calculate thereafter the theoretical photonic heat conductance in the small temperature difference limit, $G_{\gamma}=\dot{Q}_{\gamma}/(T_1-T_2)$. With $T_{\mathrm{m}}=(T_1+T_2)/2$ the mean temperature and from the heat flow expression (\ref{heat_Landauer}), we obtain 

\begin{multline} 
\label{Conductance_theo_new}
G_{\gamma}(n_{\mathrm{g}},T_{\mathrm{m}})=\frac{2k_{\mathrm{B}}^2T_{\mathrm{m}}R_1R_2}{\pi\hbar(R_1+R_2)^2} \\ \times\int_{0}^{\infty}\frac{x^2e^x\mathrm{d}x}{(e^x-1)^2}\frac{1}{1+x^2/x_{\mathrm{c}}^2(n_{\mathrm{g}})}.
\end{multline}

Here $x_{\mathrm{c}}(n_{\mathrm{g}})=\hbar\omega_{\mathrm{c}}(n_{\mathrm{g}})/k_{\mathrm{B}}T_{\mathrm{m}}$ is the reduced circuit cut-off frequency. Incidentally, in our mismatched situation this parameter introduces an additional dependence in temperature, leading to a departure from the simple $G_{\gamma}\propto T_{\mathrm{m}}$ picture \cite{Meschke2006}. In Fig. \ref{fig3}a) we see that despite its simplicity, our model reproduces well the main features of our experimental data, with essentially no free parameters. The thermal conductances at odd and even $n_{\mathrm{g}}$ are shown in Fig. \ref{fig3}b), for two cryostat temperatures $T_0=100$ and 150 mK, as a function of the mean electron temperature $T_{\mathrm{m}}$, again in good agreement with the model. A refined model may include e.g. anharmonicity, phase diffusion, the junction asymmetry, a finite stray inductance, as well as quasiparticle poisoning \cite{Aumentado2004,Ferguson2006} (see Supplementary Discussion). In addition, for large source temperatures, the gradient becomes large enough for the thermal conductance to be ill-defined.

\section{Discussion}
The performance of the device, which encompasses both the contrast in Coulomb oscillations and the maximum achieved heat flow, may be condensed in the coefficient $\beta=\mathcal{C}\times G_{\gamma,\mathrm{max}}/G_{\mathrm{Q}}$, where 1 indicates maximal performance. A device with $\beta=1$ would have a low-pass cut-off frequency $\omega_{\mathrm{c}}\gg k_{\mathrm{B}}T/\hbar$ in open position (fully quantum-limited heat conduction) and $\omega_{\mathrm{c}}\ll k_{\mathrm{B}}T/\hbar$ in closed position (fully filtered thermal noise). $\beta$ is rather small in our experiment (at best $0.1$) but may be improved with optimized device parameters for e.g. microbolometry \cite{Kokkoniemi2019,Anghel2020,Anghel2020prb} or refrigeration \cite{Tan2017} purposes. For instance, the ratio $E_{\mathrm{J}}/E_{\mathrm{c}}$ which determines $G_{\gamma,\mathrm{min}}$ may be reduced by decreasing the junctions size, while the negative impact of the subsequently reduced $E_{\mathrm{J}}$ on $G_{\gamma,\mathrm{max}}$ may be compensated by designing more resistive metallic baths in order to improve impedance matching, as long as $R_{1,2}\ll R_{\mathrm{Q}}$.

Our experiment establishes that the electron-photon relaxation mechanism can be controlled with electric field, down to a single electric charge level, in a dual manner to magnetic field control down to a single flux quantum. This could allow for instance sensitive thermal charge detection with minimal back-action from a temperature (rather than voltage) biased electrometer. Note that the recent demonstration of reversible gate-controlled suppression of supercurrent in conventional superconducting nano-constrictions \cite{DeSimoni2018,ritter2020superconducting,alegria2020highenergy} may be conveniently employed for photonic heat control as well. Here, despite the fundamentally different possible microscopic mechanisms at work, the constriction could advantageously replace the CPT, reducing the fabrication complexity. More fundamentally, a natural development would be to associate magnetic and electric field control on the same circuit to explore many-body effects due to e.g. a high-impedance environment \cite{Saito2013,PuertasMartinez2019} on heat transport, which should lead to non-trivial thermal conductance laws. In addition, the finite frequency content of noise exchange between the resistors is not addressed: a suitable setup would allow to monitor nonequilibrium voltage fluctuations \cite{Ciliberto2013} at both resistors' ends. This could extend to the quantum regime investigations of entropy production by a heat flow \cite{Ciliberto2013,Golubev2015statistics}.

\section{Methods}
\textit{Fabrication and setup.} All the junctions, contacts and leads were fabricated in a single electron-beam lithography step using the Dolan bridge technique \cite{Dolan1977}. A silicon wafer with 100 nm grown silicon oxide was coated with a stack of a $1~\mu$m thick layer of poly(methylmetalcrylate-methacrylic) acid P(MMA-MAA) resist spun for 60 s at 4000 rpm and baked at $160~^\circ$C in three steps and on top of it a 100 nm thick layer of polymethyl-metacrylate (PMMA) spun for 60 s at 4000 rpm and baked at $160~^\circ$C. The samples were patterned with electron beam lithography and subsequently developed using methyl isobutyl ketone (MIBK 1:3 Isopropanol) for PMMA and methylglycol methanol (1:2) to create the undercut in the MAA resist. The metallic parts were evaporated in three steps in the following order: Al, Al, Cu, with an in-situ oxidation step under low oxygen pressure in between the two first steps to create the tunnel barriers for both NIS probes and Josephson junctions. The clean contacts necessary for a lossless transmission between the normal and superconducting parts were created through the second and third evaporation steps. The resist was then removed using hot acetone. 

The sample was mounted in a stage with double brass enclosure acting as a radiation shield. The stage was thermally anchored to the mixing chamber of a small homemade dilution refrigerator with 50 mK base temperature. All lines were filtered with standard lossy coaxial cables with bandwidth $0-\sim 10$ kHz. Amplification of the output voltage signals at the ends of the NIS probes was realized using a room temperature low noise voltage amplifier Femto DLVPA-100-F-D. The DC signals were applied and read using standard programmable sources and multimeters. The effective integration bandwidth around the oscillator frequency for AC measurements was 0.26 Hz. The calibration of the local electronic thermometers was done by monitoring the voltage drop at the ends of the current-biased ($I_{th}\approx 160~\mathrm{pA}$) SINIS configuration while ramping up the cryostat temperature up to 350 mK (for more details see e.g. Ref. \cite{Giazotto2006}).

\textit{Thermal conductance measurement.} A precise observation of the heat flow modulation and averaging of even a moderate number of datasets is made difficult in a pure DC measurement of the electron temperature by unavoidable charge noise that manifests in single electron devices \cite{Zimmerli1992} when using long measurement times. Nevertheless, the DC values are recorded as a reference throughout the gate sweep with a typical uncertainty of $\pm$1 mK for a $\sim 1$ s averaging time, which is too large for a straight DC measurement where gate modulation depths are of this order but very small when measuring the thermal conductance with the lock-in technique (see below).

The heat balance equations are written in the main text. To measure the heat conductance we impose a small, AC heating signal on top of the DC one that establishes the thermal gradient. The AC frequency $f\sim 77$ Hz is small enough, on the one hand, for the quasi-equilibrium temperature of the electron gas to be defined (the electron-electron scattering time $\tau_{\mathrm{e}-\mathrm{e}}\sim 1$ ns $\ll f^{-1}$ \cite{Pothier1997}), and for the steady-state hypothesis to be valid at any relevant measurement timescale on the other hand: indeed, the typical energy relaxation timescale is upper bounded by the electron-phonon relaxation time $\tau_{\mathrm{e-ph}}\sim 10-100~\mu$s at 100 mK in copper \cite{Roukes1985}, which is much shorter than the typical AC modulation timescale $f^{-1}$. Therefore the power balance equation written in the main text can be re-written for a steady state displaced from $(T_1,T_2)$ to $(T_1+\delta T_1,T_2+\delta T_2)$, and expanded at first order in the increments $\delta T_{1,2}\ll T_{1,2}$, assuming the phonon temperature $T_0$ (taken equal to the cryostat temperature), remains constant:
\begin{multline}
\label{heat_balance_drain_DL}
\dot{Q}_{\mathrm{ext},2}\approx\dot{Q}_{\mathrm{e-ph},2}(T_2,T_0)+5\Sigma\Omega T_2^4\delta T_2\\
-\left[\dot{Q}_{\gamma}(T_1,T_2)+\frac{\partial \dot{Q}_{\gamma}}{\partial T_1}\bigg|_{T_1}\delta T_1+\frac{\partial \dot{Q}_{\gamma}}{\partial T_2}\bigg|_{T_2}\delta T_2\right].
\end{multline}
There we identify the power balance terms for steady-state $(T_1,T_2)$ which cancel out. Noticing that $T_1-T_2\ll T_{1,2,\mathrm{m}}$, and disregarding thermal rectification phenomena \cite{Senior2020} (the couplings to the resistances are nominally identical), we can make the following approximation that defines the experimental thermal conductance at the mean temperature $T_{\mathrm{m}}$:
\begin{equation}
\label{therm_cond_exp_approx}
G_{\gamma}(T_{\mathrm{m}})\equiv\frac{\partial \dot{Q}_{\gamma}}{\partial T_2}\bigg|_{T_2\rightarrow T_{\mathrm{m}}}\approx-\frac{\partial \dot{Q}_{\gamma}}{\partial T_1}\bigg|_{T_1\rightarrow T_{\mathrm{m}}},
\end{equation}
with corrections up to a factor $(T_1-T_2)/2T_{\mathrm{m}}$, which become important for gradients larger than $\sim 50$ mK, limiting the applicability of the method to roughly a source temperature of 230 mK. Thanks to the linearity, under these conditions, of Eq. (\ref{heat_balance_drain_DL}), we can replace the increments by their RMS value measured with a lock-in amplifier. Keeping for them the same notation and rearranging the terms in Eq. (\ref{heat_balance_drain_DL}), we finally obtain the value of thermal conductance extracted from lock-in measurements and used in the main text. Note that given our low modulation frequency we expect and indeed observe a negligible quadrature response of the lock-in amplifier read-out for a $0^{\circ}$ phase reference. Such a response should be significant only at AC heating frequencies comparable with or higher than the electron-photon or electron-phonon relaxation rates \cite{Pinsolle2016} (hence at kHz frequencies or above), but may also be visible at lower frequencies due to spurious capacitive cross-talk in the AC line which increases upon increasing the heating signal frequency.

\section{Acknowledgements}
We thank E. Aurell, F. Blanchet, A. Gubaydullin, E. Mannila, Yu. Pashkin and K. Saito for helpful discussions. This work was performed as part of the Academy of Finland Centre of Excellence program (project 310257). We acknowledge the provision of facilities by Aalto University at OtaNano-Micronova Nanofabrication Centre.

\bibliography{Phot_HeatCB_paper}

\begin{thebibliography}{10}

\bibitem{Gely2019}
M.~F. Gely, M.~Kounalakis, C.~Dickel, J.~Dalle, R.~Vatr{\'e}, B.~Baker, M.~D.
  Jenkins, and G.~A. Steele, ``Observation and stabilization of photonic fock
  states in a hot radio-frequency resonator,'' {\em Science}, vol.~363,
  no.~6431, pp.~1072--1075, 2019.

\bibitem{Giazotto2012}
F.~Giazotto and M.~J. Mart{\'i}nez-P{\'e}rez, ``The josephson heat
  interferometer,'' {\em Nature}, vol.~492, pp.~401--405 --, Dec 2012.

\bibitem{Pepa2014}
M.~Jos{\'e} Mart{\'i}nez-P{\'e}rez and F.~Giazotto, ``A quantum diffractor for
  thermal flux,'' {\em Nature Communications}, vol.~5, p.~3579, Apr 2014.

\bibitem{Pendry_1983}
J.~B. Pendry, ``Quantum limits to the flow of information and entropy,'' {\em
  Journal of Physics A: Mathematical and General}, vol.~16, pp.~2161--2171, jul
  1983.

\bibitem{Schwab2000}
K.~Schwab, E.~A. Henriksen, J.~M. Worlock, and M.~L. Roukes, ``Measurement of
  the quantum of thermal conductance,'' {\em Nature}, vol.~404, no.~6781,
  pp.~974--977, 2000.

\bibitem{Meschke2006}
M.~Meschke, W.~Guichard, and J.~P. Pekola, ``Single-mode heat conduction by
  photons,'' {\em Nature}, vol.~444, no.~7116, pp.~187--190, 2006.

\bibitem{Chiatti2006}
O.~Chiatti, J.~T. Nicholls, Y.~Y. Proskuryakov, N.~Lumpkin, I.~Farrer, and
  D.~A. Ritchie, ``Quantum thermal conductance of electrons in a
  one-dimensional wire,'' {\em Phys. Rev. Lett.}, vol.~97, p.~056601, Aug 2006.

\bibitem{Jezouin2013}
S.~Jezouin, F.~D. Parmentier, A.~Anthore, U.~Gennser, A.~Cavanna, Y.~Jin, and
  F.~Pierre, ``Quantum limit of heat flow across a single electronic channel,''
  {\em Science}, vol.~342, no.~6158, pp.~601--604, 2013.

\bibitem{Partanen2016}
M.~Partanen, K.~Y. Tan, J.~Govenius, R.~E. Lake, M.~K. M{\"a}kel{\"a},
  T.~Tanttu, and M.~M{\"o}tt{\"o}nen, ``Quantum-limited heat conduction over
  macroscopic distances,'' {\em Nature Physics}, vol.~12, no.~5, pp.~460--464,
  2016.

\bibitem{Banerjee2017}
M.~Banerjee, M.~Heiblum, A.~Rosenblatt, Y.~Oreg, D.~E. Feldman, A.~Stern, and
  V.~Umansky, ``Observed quantization of anyonic heat flow,'' {\em Nature},
  vol.~545, pp.~75--79, Apr 2017.

\bibitem{Cui1192}
L.~Cui, W.~Jeong, S.~Hur, M.~Matt, J.~C. Kl{\"o}ckner, F.~Pauly, P.~Nielaba,
  J.~C. Cuevas, E.~Meyhofer, and P.~Reddy, ``Quantized thermal transport in
  single-atom junctions,'' {\em Science}, vol.~355, no.~6330, pp.~1192--1195,
  2017.

\bibitem{Saito2013}
K.~Saito and T.~Kato, ``Kondo signature in heat transfer via a local two-state
  system,'' {\em Phys. Rev. Lett.}, vol.~111, p.~214301, Nov 2013.

\bibitem{Dutta2017}
B.~Dutta, J.~T. Peltonen, D.~S. Antonenko, M.~Meschke, M.~A. Skvortsov,
  B.~Kubala, J.~K\"onig, C.~B. Winkelmann, H.~Courtois, and J.~P. Pekola,
  ``Thermal conductance of a single-electron transistor,'' {\em Phys. Rev.
  Lett.}, vol.~119, p.~077701, Aug 2017.

\bibitem{Sivre2017}
E.~Sivre, A.~Anthore, F.~D. Parmentier, A.~Cavanna, U.~Gennser, A.~Ouerghi,
  Y.~Jin, and F.~Pierre, ``Heat coulomb blockade of one ballistic channel,''
  {\em Nature Physics}, vol.~14, pp.~145--148, Oct 2017.

\bibitem{Sivre2019}
E.~Sivre, H.~Duprez, A.~Anthore, A.~Aassime, F.~D. Parmentier, A.~Cavanna,
  A.~Ouerghi, U.~Gennser, and F.~Pierre, ``Electronic heat flow and thermal
  shot noise in quantum circuits,'' {\em Nature Communications}, vol.~10,
  no.~1, p.~5638, 2019.

\bibitem{Ronzani2018}
A.~Ronzani, B.~Karimi, J.~Senior, Y.-C. Chang, J.~T. Peltonen, C.~Chen, and
  J.~P. Pekola, ``Tunable photonic heat transport in a quantum heat valve,''
  {\em Nature Physics}, vol.~14, no.~10, pp.~991--995, 2018.

\bibitem{Tan2017}
K.~Y. Tan, M.~Partanen, R.~E. Lake, J.~Govenius, S.~Masuda, and
  M.~M{\"o}tt{\"o}nen, ``Quantum-circuit refrigerator,'' {\em Nature
  Communications}, vol.~8, no.~1, p.~15189, 2017.

\bibitem{Partanen2018}
M.~Partanen, K.~Y. Tan, S.~Masuda, J.~Govenius, R.~E. Lake, M.~Jenei,
  L.~Gr{\"o}nberg, J.~Hassel, S.~Simbierowicz, V.~Vesterinen, J.~Tuorila,
  T.~Ala-Nissila, and M.~M{\"o}tt{\"o}nen, ``Flux-tunable heat sink for quantum
  electric circuits,'' {\em Scientific Reports}, vol.~8, no.~1, p.~6325, 2018.

\bibitem{Kokkoniemi2019}
R.~Kokkoniemi, J.~Govenius, V.~Vesterinen, R.~E. Lake, A.~M. Gunyh{\'o}, K.~Y.
  Tan, S.~Simbierowicz, L.~Gr{\"o}nberg, J.~Lehtinen, M.~Prunnila, J.~Hassel,
  A.~Lamminen, O.-P. Saira, and M.~M{\"o}tt{\"o}nen, ``Nanobolometer with
  ultralow noise equivalent power,'' {\em Communications Physics}, vol.~2,
  p.~124, Oct 2019.

\bibitem{Anghel2020}
D.~Anghel and L.~Kuzmin, ``Cold-electron bolometer as a 1-cm-wavelength photon
  counter,'' {\em Phys. Rev. Applied}, vol.~13, p.~024028, Feb 2020.

\bibitem{Anghel2020prb}
D.~V. Anghel, K.~Kulikov, Y.~M. Galperin, and L.~S. Kuzmin, ``Electromagnetic
  radiation detectors based on josephson junctions: Effective hamiltonian,''
  {\em Phys. Rev. B}, vol.~101, p.~024511, Jan 2020.

\bibitem{Johnson1928}
J.~B. Johnson, ``Thermal agitation of electricity in conductors,'' {\em Phys.
  Rev.}, vol.~32, pp.~97--109, Jul 1928.

\bibitem{Nyquist1928}
H.~Nyquist, ``Thermal agitation of electric charge in conductors,'' {\em Phys.
  Rev.}, vol.~32, pp.~110--113, Jul 1928.

\bibitem{Schmidt2004}
D.~R. Schmidt, R.~J. Schoelkopf, and A.~N. Cleland, ``Photon-mediated thermal
  relaxation of electrons in nanostructures,'' {\em Phys. Rev. Lett.}, vol.~93,
  p.~045901, Jul 2004.

\bibitem{Pascal2011}
L.~M.~A. Pascal, H.~Courtois, and F.~W.~J. Hekking, ``Circuit approach to
  photonic heat transport,'' {\em Phys. Rev. B}, vol.~83, p.~125113, Mar 2011.

\bibitem{Timofeev2009}
A.~V. Timofeev, M.~Helle, M.~Meschke, M.~M\"ott\"onen, and J.~P. Pekola,
  ``Electronic refrigeration at the quantum limit,'' {\em Phys. Rev. Lett.},
  vol.~102, p.~200801, May 2009.

\bibitem{Blencowe1999}
M.~P. Blencowe, ``Quantum energy flow in mesoscopic dielectric structures,''
  {\em Phys. Rev. B}, vol.~59, pp.~4992--4998, Feb 1999.

\bibitem{Ojanen2008}
T.~Ojanen and A.-P. Jauho, ``Mesoscopic photon heat transistor,'' {\em Phys.
  Rev. Lett.}, vol.~100, p.~155902, Apr 2008.

\bibitem{Scheibner2005}
R.~Scheibner, H.~Buhmann, D.~Reuter, M.~N. Kiselev, and L.~W. Molenkamp,
  ``Thermopower of a kondo spin-correlated quantum dot,'' {\em Phys. Rev.
  Lett.}, vol.~95, p.~176602, Oct 2005.

\bibitem{Thierschmann2015}
H.~Thierschmann, R.~S{\'a}nchez, B.~Sothmann, F.~Arnold, C.~Heyn, W.~Hansen,
  H.~Buhmann, and L.~W. Molenkamp, ``Three-terminal energy harvester with
  coupled quantum dots,'' {\em Nature Nanotechnology}, vol.~10, pp.~854--858,
  Oct 2015.

\bibitem{Josefsson2018}
M.~Josefsson, A.~Svilans, A.~M. Burke, E.~A. Hoffmann, S.~Fahlvik,
  C.~Thelander, M.~Leijnse, and H.~Linke, ``A quantum-dot heat engine operating
  close to the thermodynamic efficiency limits,'' {\em Nature Nanotechnology},
  vol.~13, pp.~920--924, Oct 2018.

\bibitem{Dutta2019}
B.~Dutta, D.~Majidi, A.~Garc{\'i}a~Corral, P.~A. Erdman, S.~Florens, T.~A.
  Costi, H.~Courtois, and C.~B. Winkelmann, ``Direct probe of the seebeck
  coefficient in a kondo-correlated single-quantum-dot transistor,'' {\em Nano
  Letters}, vol.~19, pp.~506--511, Jan 2019.

\bibitem{Jaliel2019}
G.~Jaliel, R.~K. Puddy, R.~S\'anchez, A.~N. Jordan, B.~Sothmann, I.~Farrer,
  J.~P. Griffiths, D.~A. Ritchie, and C.~G. Smith, ``Experimental realization
  of a quantum dot energy harvester,'' {\em Phys. Rev. Lett.}, vol.~123,
  p.~117701, Sep 2019.

\bibitem{Dutta2020}
B.~Dutta, D.~Majidi, N.~W. Talarico, N.~L. Gullo, C.~B. Winkelmann, and
  H.~Courtois, ``A single-quantum-dot heat valve,'' 2020.

\bibitem{Molenkamp1992}
L.~W. Molenkamp, T.~Gravier, H.~van Houten, O.~J.~A. Buijk, M.~A.~A. Mabesoone,
  and C.~T. Foxon, ``Peltier coefficient and thermal conductance of a quantum
  point contact,'' {\em Phys. Rev. Lett.}, vol.~68, pp.~3765--3768, Jun 1992.

\bibitem{joyez1995}
P.~Joyez, {\em Le transistor \`a une paire de Cooper : un syst\`eme quantique
  macroscopique}.
\newblock PhD thesis, 1995.
\newblock Universit\'{e} Paris 6.

\bibitem{Andreev1964}
A.~Andreev, ``Thermal conductivity of the intermediate state of
  superconductors,'' {\em Zh. Eksperim. i Teor. Fiz.}, vol.~19, May 1964.

\bibitem{Giazotto2006}
F.~Giazotto, T.~T. Heikkil\"a, A.~Luukanen, A.~M. Savin, and J.~P. Pekola,
  ``Opportunities for mesoscopics in thermometry and refrigeration: Physics and
  applications,'' {\em Rev. Mod. Phys.}, vol.~78, pp.~217--274, Mar 2006.

\bibitem{Pothier1997}
H.~Pothier, S.~Gu\'eron, N.~O. Birge, D.~Esteve, and M.~H. Devoret, ``Energy
  distribution function of quasiparticles in mesoscopic wires,'' {\em Phys.
  Rev. Lett.}, vol.~79, pp.~3490--3493, Nov 1997.

\bibitem{Pinsolle2016}
E.~Pinsolle, A.~Rousseau, C.~Lupien, and B.~Reulet, ``Direct measurement of the
  electron energy relaxation dynamics in metallic wires,'' {\em Phys. Rev.
  Lett.}, vol.~116, p.~236601, Jun 2016.

\bibitem{Aumentado2004}
J.~Aumentado, M.~W. Keller, J.~M. Martinis, and M.~H. Devoret, ``Nonequilibrium
  quasiparticles and $2e$ periodicity in single-cooper-pair transistors,'' {\em
  Phys. Rev. Lett.}, vol.~92, p.~066802, Feb 2004.

\bibitem{Devoret1997}
M.~H. Devoret, {\em Quantum fluctuations in electrical circuits}.
\newblock France: Edition de Physique, 1997.

\bibitem{SuppMat}
``See supplementary information for detailed calculations and modelling.''

\bibitem{Ferguson2006}
A.~J. Ferguson, N.~A. Court, F.~E. Hudson, and R.~G. Clark, ``Microsecond
  resolution of quasiparticle tunneling in the single-cooper-pair transistor,''
  {\em Phys. Rev. Lett.}, vol.~97, p.~106603, Sep 2006.

\bibitem{DeSimoni2018}
G.~De~Simoni, F.~Paolucci, P.~Solinas, E.~Strambini, and F.~Giazotto,
  ``Metallic supercurrent field-effect transistor,'' {\em Nature
  Nanotechnology}, vol.~13, pp.~802--805, Sep 2018.

\bibitem{ritter2020superconducting}
M.~F. Ritter, A.~Fuhrer, D.~Z. Haxell, S.~Hart, P.~Gumann, H.~Riel, and
  F.~Nichele, ``A superconducting switch actuated by injection of high energy
  electrons,'' 2020.

\bibitem{alegria2020highenergy}
L.~D. Alegria, C.~G. Bøttcher, A.~K. Saydjari, A.~T. Pierce, S.~H. Lee, S.~P.
  Harvey, U.~Vool, and A.~Yacoby, ``High-energy quasiparticle injection in
  mesoscopic superconductors,'' 2020.

\bibitem{PuertasMartinez2019}
J.~Puertas~Mart{\'i}nez, S.~L{\'e}ger, N.~Gheeraert, R.~Dassonneville,
  L.~Planat, F.~Foroughi, Y.~Krupko, O.~Buisson, C.~Naud, W.~Hasch-Guichard,
  S.~Florens, I.~Snyman, and N.~Roch, ``A tunable josephson platform to explore
  many-body quantum optics in circuit-qed,'' {\em npj Quantum Information},
  vol.~5, no.~1, p.~19, 2019.

\bibitem{Ciliberto2013}
S.~Ciliberto, A.~Imparato, A.~Naert, and M.~Tanase, ``Heat flux and entropy
  produced by thermal fluctuations,'' {\em Phys. Rev. Lett.}, vol.~110,
  p.~180601, Apr 2013.

\bibitem{Golubev2015statistics}
D.~S. Golubev and J.~P. Pekola, ``Statistics of heat exchange between two
  resistors,'' {\em Phys. Rev. B}, vol.~92, p.~085412, Aug 2015.

\bibitem{Dolan1977}
G.~J. Dolan, ``Offset masks for lift‐off photoprocessing,'' {\em Applied
  Physics Letters}, vol.~31, no.~5, pp.~337--339, 1977.

\bibitem{Zimmerli1992}
G.~Zimmerli, T.~M. Eiles, R.~L. Kautz, and J.~M. Martinis, ``Noise in the
  coulomb blockade electrometer,'' {\em Applied Physics Letters}, vol.~61,
  no.~2, pp.~237--239, 1992.

\bibitem{Roukes1985}
M.~L. Roukes, M.~R. Freeman, R.~S. Germain, R.~C. Richardson, and M.~B.
  Ketchen, ``Hot electrons and energy transport in metals at millikelvin
  temperatures,'' {\em Phys. Rev. Lett.}, vol.~55, pp.~422--425, Jul 1985.

\bibitem{Senior2020}
J.~Senior, A.~Gubaydullin, B.~Karimi, J.~T. Peltonen, J.~Ankerhold, and J.~P.
  Pekola, ``Heat rectification via a superconducting artificial atom,'' {\em
  Communications Physics}, vol.~3, no.~1, p.~40, 2020.

\bibitem{Pekola1994}
J.~P. Pekola, K.~P. Hirvi, J.~P. Kauppinen, and M.~A. Paalanen, ``Thermometry
  by arrays of tunnel junctions,'' {\em Phys. Rev. Lett.}, vol.~73,
  pp.~2903--2906, Nov 1994.

\bibitem{Leivo1996}
M.~M. Leivo, J.~P. Pekola, and D.~V. Averin, ``Efficient peltier refrigeration
  by a pair of normal metal/insulator/superconductor junctions,'' {\em Applied
  Physics Letters}, vol.~68, no.~14, pp.~1996--1998, 1996.

\bibitem{Viisanen2018}
K.~L. Viisanen and J.~P. Pekola, ``Anomalous electronic heat capacity of copper
  nanowires at sub-kelvin temperatures,'' {\em Phys. Rev. B}, vol.~97,
  p.~115422, Mar 2018.

\bibitem{Wang2019}
L.~B. Wang, O.-P. Saira, D.~S. Golubev, and J.~P. Pekola, ``Crossover between
  electron-phonon and boundary-resistance limits to thermal relaxation in
  copper films,'' {\em Phys. Rev. Applied}, vol.~12, p.~024051, Aug 2019.

\bibitem{ThesisCottet}
A.~Cottet, {\em {Impl{\'e}mentation d'un bit quantique dans un circuit
  supraconducteur / Implementation of a quantum bit in a superconducting
  circuit}}.
\newblock Theses, {Universit{\'e} Pierre et Marie Curie - Paris VI}, Sept.
  2002.

\bibitem{Stewart1968}
W.~C. Stewart, ``Current voltage characteristics of josephson junctions,'' {\em
  Applied Physics Letters}, vol.~12, no.~8, pp.~277--280, 1968.

\bibitem{Matveev1993}
K.~A. Matveev, M.~Gisself\"alt, L.~I. Glazman, M.~Jonson, and R.~I. Shekhter,
  ``Parity-induced suppression of the coulomb blockade of josephson
  tunneling,'' {\em Phys. Rev. Lett.}, vol.~70, pp.~2940--2943, May 1993.

\bibitem{tinkham2004}
M.~Tinkham, {\em Introduction to Superconductivity}.
\newblock Dover Books on Physics Series, Dover Publications, 2004.

\bibitem{Martinis2009}
J.~M. Martinis, M.~Ansmann, and J.~Aumentado, ``Energy decay in superconducting
  josephson-junction qubits from nonequilibrium quasiparticle excitations,''
  {\em Phys. Rev. Lett.}, vol.~103, p.~097002, Aug 2009.

\bibitem{Mannila2019}
E.~T. Mannila, V.~F. Maisi, H.~Q. Nguyen, C.~M. Marcus, and J.~P. Pekola,
  ``Detecting parity effect in a superconducting device in the presence of
  parity switches,'' {\em Phys. Rev. B}, vol.~100, p.~020502, Jul 2019.

\bibitem{Catelani2011}
G.~Catelani, J.~Koch, L.~Frunzio, R.~J. Schoelkopf, M.~H. Devoret, and L.~I.
  Glazman, ``Quasiparticle relaxation of superconducting qubits in the presence
  of flux,'' {\em Phys. Rev. Lett.}, vol.~106, p.~077002, Feb 2011.

\bibitem{Riste2013}
D.~Rist{\`e}, C.~C. Bultink, M.~J. Tiggelman, R.~N. Schouten, K.~W. Lehnert,
  and L.~DiCarlo, ``Millisecond charge-parity fluctuations and induced
  decoherence in a superconducting transmon qubit,'' {\em Nature
  Communications}, vol.~4, p.~1913, May 2013.

\bibitem{Joyez1994}
P.~Joyez, P.~Lafarge, A.~Filipe, D.~Esteve, and M.~H. Devoret, ``Observation of
  parity-induced suppression of josephson tunneling in the superconducting
  single electron transistor,'' {\em Phys. Rev. Lett.}, vol.~72,
  pp.~2458--2461, Apr 1994.

\bibitem{Mannik2004}
J.~M\"annik and J.~E. Lukens, ``Effect of measurement on the periodicity of the
  coulomb staircase of a superconducting box,'' {\em Phys. Rev. Lett.},
  vol.~92, p.~057004, Feb 2004.

\bibitem{barone1982physics}
A.~Barone and G.~Patern{\`o}, {\em Physics and applications of the Josephson
  effect}.
\newblock UMI Out-of-Print Books on Demand, Wiley, 1982.

\end{thebibliography}
\bibliographystyle{ieeetr}

\onecolumngrid
\appendix
\setcounter{figure}{0}
\renewcommand{\thepage}{S\arabic{page}} 
\renewcommand{\thesection}{S\arabic{section}}  
\renewcommand{\thetable}{S\arabic{table}}  
\renewcommand{\thefigure}{S\arabic{figure}} 
\section*{Supplementary Methods}
\textit{Experiment parameters.} The superconducting gap, tunnel resistances of the CPT and NIS junctions were obtained through standard current-voltage characteristic measurements. The superconducting gap, assumed equal for both Aluminium layers having the same thickness and evaporated with the same target, was measured to be $\Delta=214~\mu$eV $=k_{\mathrm{B}}\times 2.48$ K. The NIS probes tunnel resistances were found between 5 and 20 k$\Omega$, while the series connection of Josephson junctions yields a normal-state resistance $R_{\mathrm{N}}=22.5$ k$\Omega$, from which we deduce the total Josephson energy $2E_{\mathrm{J}}=E_{\mathrm{J}1}+E_{\mathrm{J}2}=k_{\mathrm{B}}\times 1.38$ K assuming identical junctions and using the Ambegaokar-Baratoff relation for a single junction $E_{\mathrm{J}}=h\Delta/4e^2R_{\mathrm{N}}$ (here each junction has a normal state resistance $R_{\mathrm{N}}/2$).

The single-electron charging energy $E_{\mathrm{c}}=e^2/2C$ of the CPT was determined by measuring the differential conductance of the system in weak Coulomb blockade regime in the normal state at $4.21$ K. In the regime $E_{\mathrm{c}}\ll k_{\mathrm{B}}T$, the zero-bias differential conductance takes a universal value \cite{Pekola1994}, $R_{\mathrm{N}}\mathrm{d}I/\mathrm{d}V|_{V=0}=1-E_{\mathrm{c}}/3k_{\mathrm{B}}T$, which yields a charging energy $E_{\mathrm{c}}=k_{\mathrm{B}}\times 0.64\pm 0.05$ K [see Supplementary Figure \ref{Peph}a)]. Note that this is the single electron charging energy, whereas in the superconducting state the relevant energy scale is $4E_{\mathrm{c}}=(2e)^2/2C$ which is bigger than $2E_{\mathrm{J}}$. However for consistency with later steps we keep the single-electron definition for $E_{\mathrm{c}}$. By measuring both switching current (see below) and output current gate modulation for voltages $V\sim 4\Delta/e$ we obtain the $1e$ gate voltage period $\Delta V_{\mathrm{g},1e}=13$ mV and deduce the gate capacitance $C_{\mathrm{g}}\approx 12$ aF.

The source and drain resistances were estimated at 4 K via four-probe measurements to be $R_i\approx 290\pm 30~\Omega$. The large uncertainty is due to the use of NIS junctions as voltage probes that measure only a fraction of the voltage drop in the resistor due to their location, as well as deviations from nominal profiles of the films and proximity superconductivity that could manifest below 1 K. The value used in the theoretical modelling of the main text is $275~\Omega$, well within the range expected.
\begin{figure}
	\includegraphics[width=17.2cm]{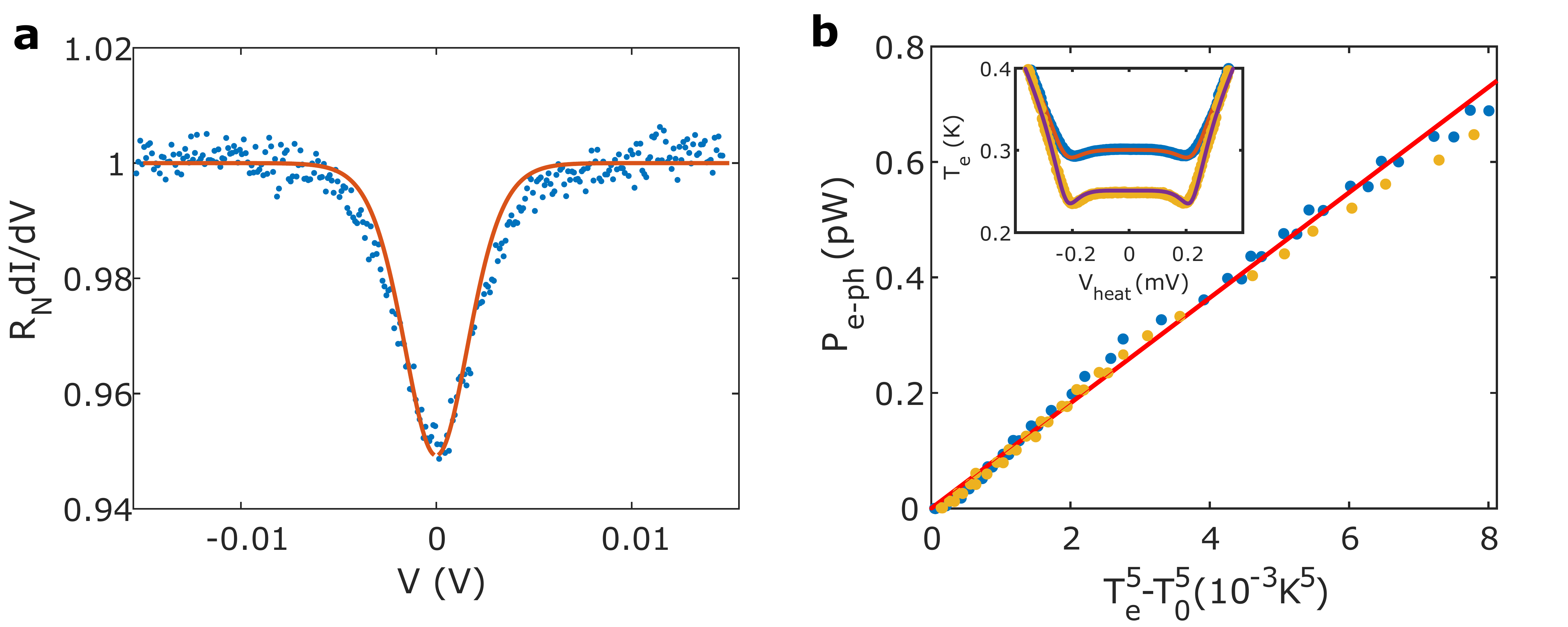}
	\caption{\textbf{Measurement of charging energy and electron-phonon coupling constant.} a) Differential conductance of the CPT in normal state at 4.21 K, normalized to the large bias ohmic value $1/R_{\mathrm{N}}$. The red curve is the application of the universal formula derived in \cite{Pekola1994} for $E_{\mathrm{c}}=k_{\mathrm{B}}\times 0.64$ K. b) Power dissipated in the source resistor through NIS heating versus the quantity $T_{\mathrm{e}}^5-T_0^5$, where $T_{\mathrm{e}}$ is the source electronic temperature, for cryostat temperatures $T_0=250$ mK (yellow dots) and 300 mK (blue dots). The fitted slope yields $\Sigma\Omega$. Inset: electron temperature in the source resistor measured with NIS thermometry versus voltage applied to another NIS junction of the source at 250 and 300 mK cryostat temperatures (same conventions). Solid lines are the application of the formula for injected power in the NIS junction (see Supplementary References \cite{Leivo1996,Giazotto2006})}.
	\label{Peph}
\end{figure}

The value $\Sigma=(3.7\pm 0.2)\times 10^9~\mathrm{W.K^{-5}.m^{-3}}$ for electron-phonon coupling quoted in the main text is slightly larger than the value commonly given in the literature ($2\times 10^9~\mathrm{W.K^{-5}.m^{-3}}$, see e.g. Supplementary References \cite{Roukes1985,Viisanen2018,Wang2019}) for thicker films but already reported previously \cite{Leivo1996} and measured independently here. As the volume $\Omega$ of the metallic piece is not known with very good accuracy, the value of $\Sigma$ may not be entirely reliable, as it depends on the value chosen for $\Omega$. We stress that this does not affect the measurements, the only experimentally relevant quantity being $\Sigma\Omega$. In Supplementary Figure \ref{Peph}b) we show the power dissipated in the resistors \cite{Leivo1996,Giazotto2006} as a function of the electronic temperature $T_e$ resulting from dissipation for two cryostat temperatures $T_0=250$ mK and 300 mK. We use such elevated bath temperatures to minimize the impact of electron-photon coupling and external heat loads.

We evaluate the parasitic external heat loads to be $\approx 2$ fW, in line with similar experimental conditions \cite{Giazotto2012}, which lead to a decoupling between electron temperatures in the resistors and the cryostat one below roughly 160 mK. This rather large temperature is well accounted by the small volume $\Omega$ of the resistors, for which the external heat load is less efficiently evacuated to the phonon bath.

\textit{Control experiment}. To confirm further that the heat flow is indeed photonic in nature, we have performed a control experiment where the loop bond closing the circuit is absent, to reduce the current fluctuations responsible for photonic heat flow \cite{Timofeev2009}, while still allowing diffusive heat transport with quasiparticles \cite{Giazotto2012}. As shown in Supplementary Figure \ref{Addfig1}, a clear reduction of the conductance is observed, down to $\approx 45$ \% of its value and oscillation contrast in the matched situation. The oscillations remain $2e$-periodic: therefore, the remaining heat flow is likely due to capacitive leakage of the connecting apparatus, leading to only partial high-pass filtering of thermal fluctuations and allowing some flow within a residual bandwidth. The observed reduction nonetheless confirms that the source-drain thermal conductance is dominated by the electron-photon coupling, that creates a remote electron-electron coupling with gate-tunable strength. 
\begin{figure}
	\includegraphics[width=8.6cm]{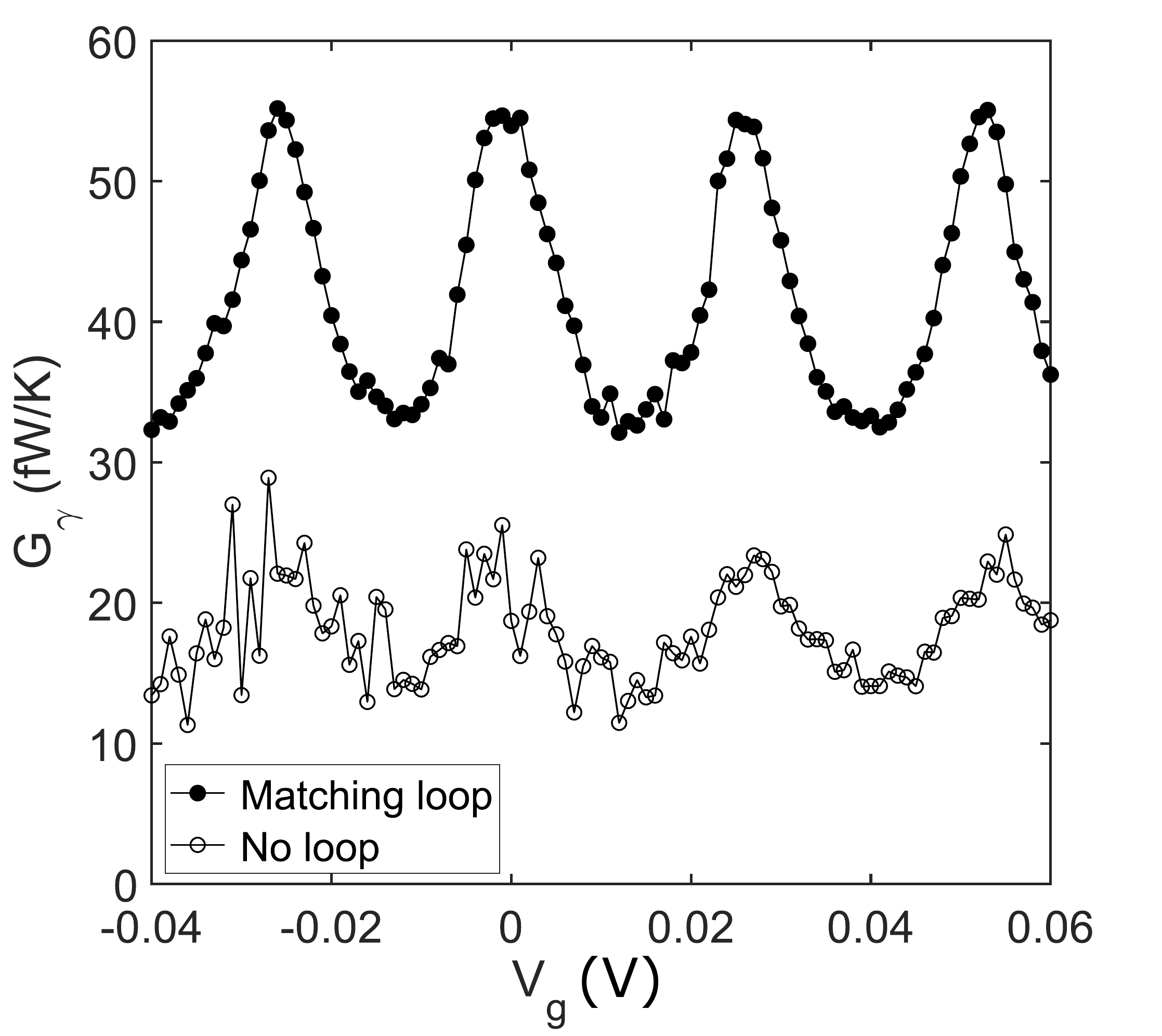}
	\caption{\textbf{Control experiment and comparison with the "loop" configuration.} Measured source-drain thermal conductance as a function of the gate voltage with (full dots) and without (empty dots) a loop-closing wire, for comparable median temperatures $T_{\mathrm{m}} = 187$ mK (loop) and 190 mK (no loop).}
	\label{Addfig1}
\end{figure}

\textit{Estimate for quasiparticle heat transfer.} The systematically observed $2e$ periodicity seem to rule out heat transport dominated by a quasiparticle diffusion mechanism, which should be $1e$ periodic. In addition, the measured subgap resistance is $R_{\mathrm{s}}\approx 2.5$ M$\Omega$. Applying Wiedemann-Franz law, which should hold at least within a numerical factor of order unity for our CPT \cite{Dutta2017}, we obtain a thermal conductance $G_{\mathrm{qp}}\approx \mathcal{L}_0T/R_{\mathrm{s}}\approx 2$ fW/K at 200 mK ($\mathcal{L}_0=2.44\times 10^{-8}$ W.$\Omega.\mathrm{K}^{-2}$ is the Lorenz number), well below the measured one both in matched and mismatched configuration. Furthermore, the electronic heat conduction is strongly attenuated along the superconducting line due to the reduced quasiparticle density \cite{Timofeev2009}, and thus the effective quasiparticle thermal conductance will be much smaller than this upper limit. This rules out for good the hypothesis of quasiparticle-mediated heat transfer. Note that this is different from quasiparticle poisoning, whose effect is to shift the induced charge by an amount $e$ after a quasiparticle tunneling event, which results in averaging the thermal conductance over its value at $n_{\mathrm{g}}$ and that at $n_{\mathrm{g}}+1$ (see below).

\section*{Supplementary Note 1}
\textit{CPT Hamiltonian and critical current derivation.} The Hamiltonian of the Cooper-pair transistor under zero voltage bias, assuming its two JJ are identical and neglecting quasiparticle excitations, writes:
\begin{equation}
\label{Hamiltonian_line}
\hat{\mathcal{H}}=E_{\mathrm{c}}(\hat{n}-n_{\mathrm{g}})^2-2E_{\mathrm{J}}\cos\hat{\phi}\cos\delta/2,
\end{equation}
where $\hat{n}$ is the number of extra paired electrons in the island, $n_{\mathrm{g}}=C_{\mathrm{g}}V_{\mathrm{g}}/e$ is the gate charge, $\delta=\delta_1+\delta_2$ is the total phase across the series connection of JJ and $\hat{\phi}=\delta_2-\delta_1$ is the phase of the island. We derive all properties over the interval $n_{\mathrm{g}}\in[-1,1]$, knowing that the energy bands are periodic in $2e$. We choose for numerical and theoretical purposes to restrict the charge state basis to the subset $\{|n=-2\rangle,|n=0\rangle,|n=2\rangle\}$. The Hamiltonian of Supplementary Equation \ref{Hamiltonian_line} writes in matrix form in the chosen basis: 
\begin{equation}
\label{Hamiltonian_matrix}
\hat{\mathcal{H}}=\begin{pmatrix}
E_{\mathrm{c}}(2+n_{\mathrm{g}})^2      & -E_{\mathrm{J}}\cos\delta/2 & 0 \\ 
-E_{\mathrm{J}}\cos\delta/2 & E_{\mathrm{c}}n_{\mathrm{g}}^2 & -E_{\mathrm{J}}\cos\delta/2 \\ 
0      & -E_{\mathrm{J}}\cos\delta/2 & E_{\mathrm{c}}(2-n_{\mathrm{g}})^2 
\end{pmatrix}.
\end{equation}
\begin{figure}
	\includegraphics[width=8.6cm]{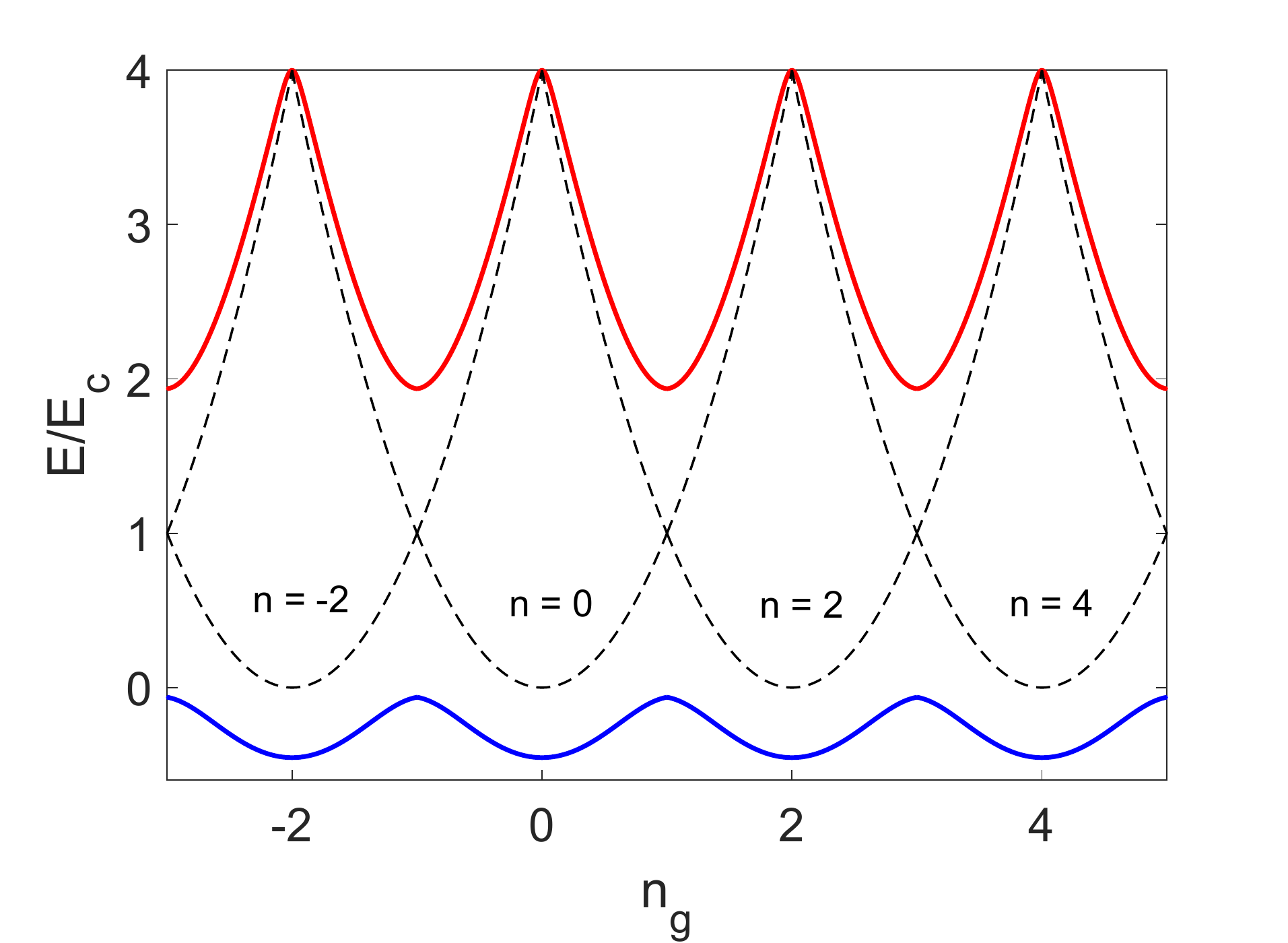}
	\includegraphics[width=8.6cm]{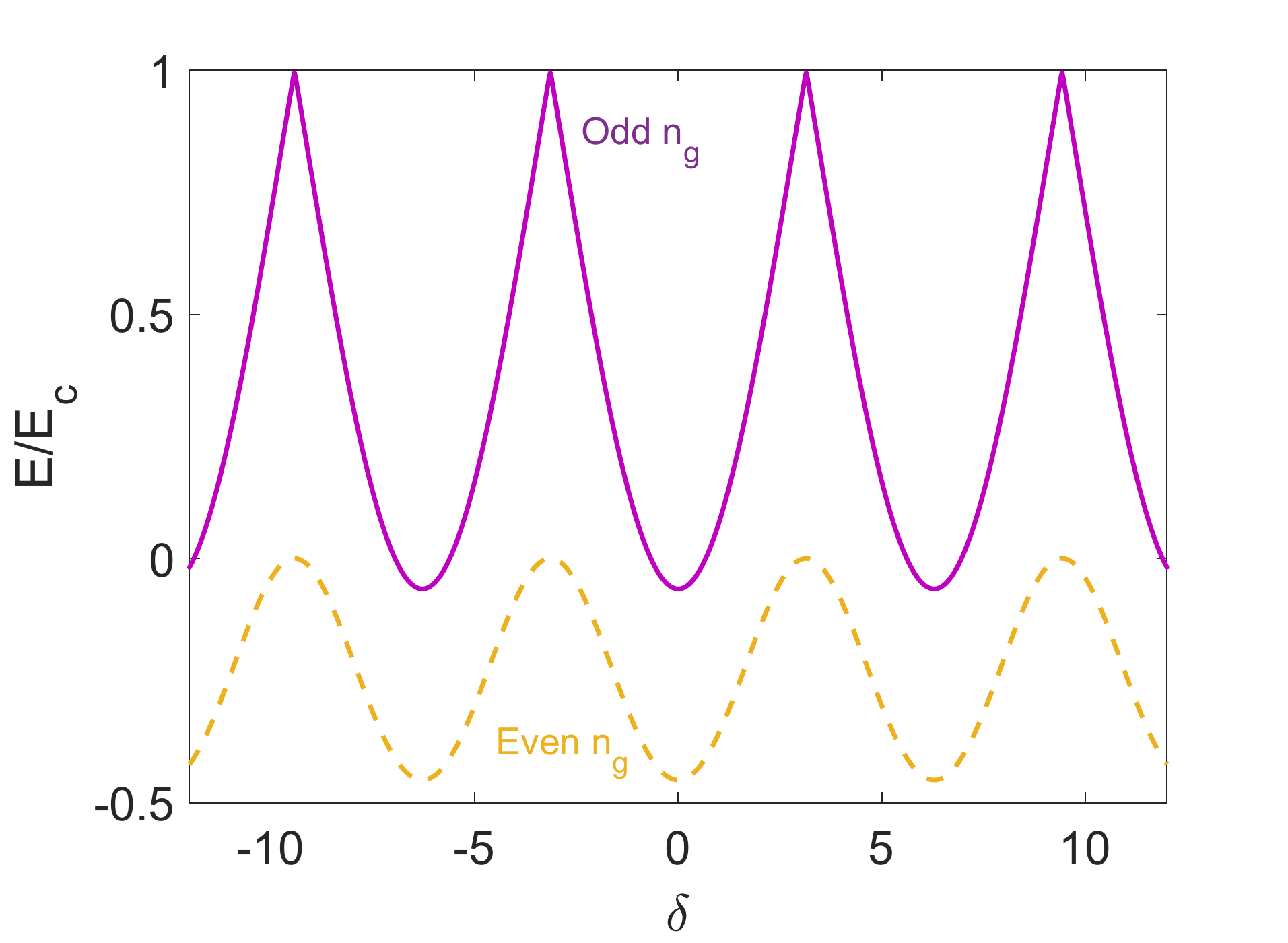}
	\caption{\textbf{Low energy configuration of the CPT.} a) Ground and first excited energy levels $|E_{0}\rangle,|E_{1}\rangle$ of the CPT as a function of the gate charge $n_{\mathrm{g}}$ for $\delta=0$ (the dotted line corresponds to bare charge states) b) Ground band as a function of the total phase $\delta$ accross the CPT for $n_{\mathrm{g}} = 0$ (yellow) and $1$ (purple) mod 2.}
	\label{Elevels}
\end{figure}

While the full Hamiltonian possesses analytical eigenenergies based on Mathieu functions \cite{ThesisCottet}, those are not easy to handle for our analysis. One can show \cite{joyez1995} that the restriction to the three lowest charge states provides fairly convenient, analytical expressions for the three first eigenenergies $E_m,m=0,1,2$ (Bloch bands) that are still good approximations of the full theory for $E_{\mathrm{c}}\sim E_{\mathrm{J}}$. After diagonalization of the truncated Hamiltonian presented in Supplementary Equation \ref{Hamiltonian_matrix}, the eigenenergies are:
\begin{equation}
\label{Bloch_bands}
E_m(n_{\mathrm{g}},\delta)=\left(\frac{8}{3}+n_{\mathrm{g}}^2\right)E_{\mathrm{c}}+\sqrt{4\lambda(\delta)}\cos\left[\frac{\theta(n_{\mathrm{g}},\delta)+2\pi(m+1)}{3}\right],
\end{equation}
where: 
\begin{equation}
\label{Bloch_bands_coeff_lambda}
\lambda(\delta)=\frac{2E_{\mathrm{J}}}{3}\cos^2\left(\frac{\delta}{2}\right)\left(E_{\mathrm{J}}+8E_{\mathrm{c}}\right),
\end{equation}
and
\begin{equation}
\theta(n_{\mathrm{g}},\delta)=\arccos\left(-\frac{4E_{\mathrm{c}}E_{\mathrm{J}}^2\cos^2\left(\delta/2\right)+64E_{\mathrm{c}}^3(1/9-n_{\mathrm{g}}^2)}{3\lambda(\delta)^{3/2}}\right).
\end{equation}
A representation of the eigenenergy bands along some chosen axes is shown in Supplementary Figure \ref{Elevels}. In particular, the charging energy term changes the shape of the effective Josephson potential represented by the ground band, as well as its barrier height.
\begin{figure}
	\centering
	\includegraphics[width=8.6cm]{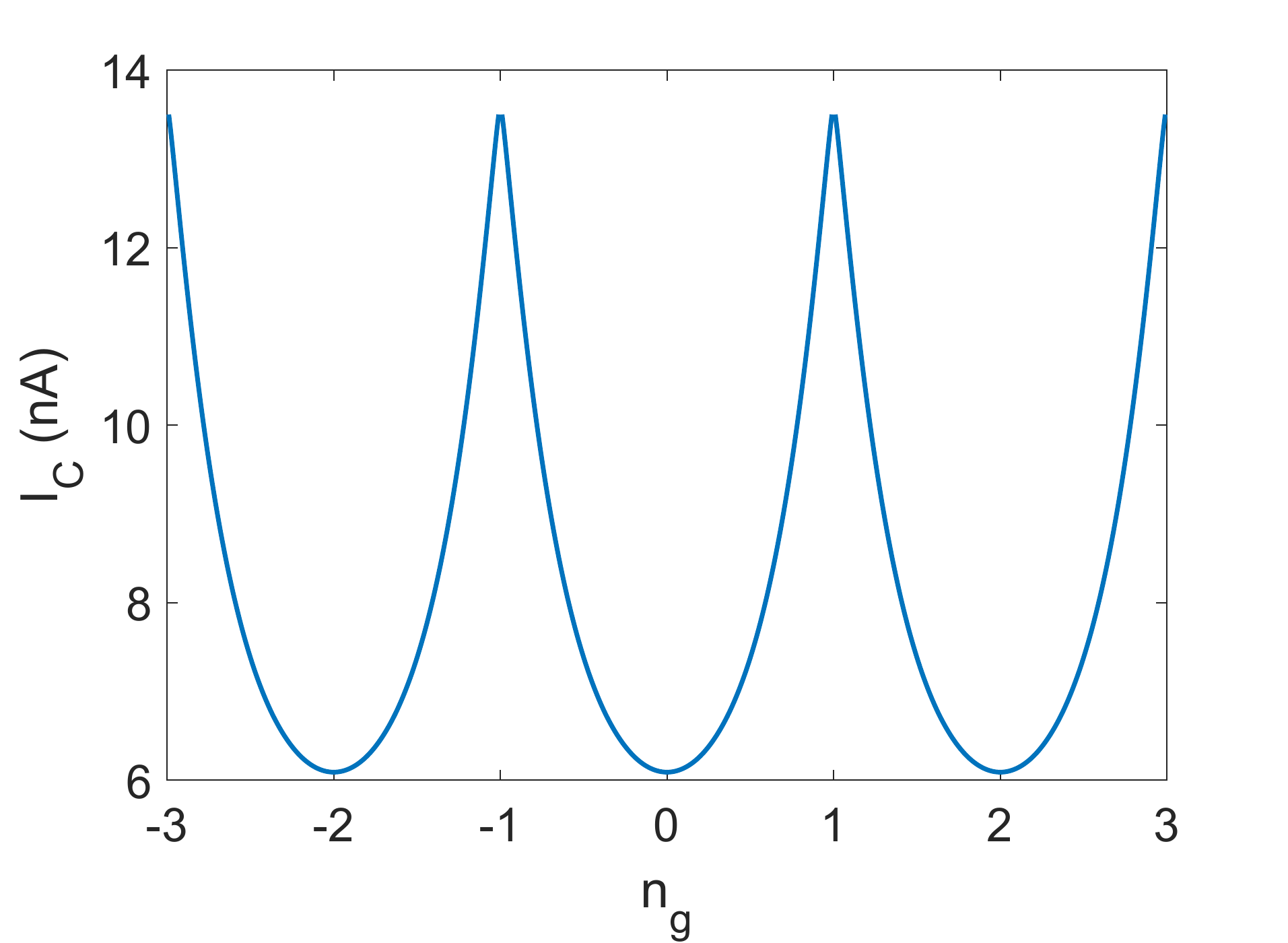}
	\caption{\textbf{Critical current.} Theoretical critical current as a function of reduced gate voltage, obtained using Supplementary Equation \ref{critical_current}.}
	\label{Ic_fig}
\end{figure}

From then on, the critical current can be derived from the supercurrent operator $\hat{I}=(2e/\hbar)\partial\hat{\mathcal{H}}/\partial\delta$:
\begin{equation}
\label{critical_current}
I_{\mathrm{C}}(n_{\mathrm{g}})=\frac{2e}{\hbar}\max_{\delta}\left\langle\frac{\partial\hat{\mathcal{H}}}{\partial\delta}\right\rangle\approx\frac{2e}{\hbar}\max_{\delta}\frac{\partial E_0(n_{\mathrm{g}},\delta)}{\partial\delta}.
\end{equation}

Here the brackets denote thermal average. For $E_{\mathrm{J}}\gtrsim k_{\mathrm{B}}T$, one can ignore the second excited state $|E_2\rangle$, but for $T\approx 100-200$ mK, there actually is a substantial probability to thermally populate the first excited state $|E_1\rangle$. However, for $E_{\mathrm{c}}\sim E_{\mathrm{J}}$ this is true only around the charge degeneracy points, i.e. for odd values of $n_{\mathrm{g}}$, where the energy splitting $E_1-E_0$ is minimum. For odd gate charge values, the maximum critical current in the ground or the excited state is the same for our symmetric model, and differs weakly for even moderate junction asymmetry, hence the approximation made in the second equality. Using the Ambegaokar-Baratoff relation for $E_{\mathrm{J}}$, the theoretical critical current can thus be readily obtained, as shown in Supplementary Figure \ref{Ic_fig}. 

The calculated critical current is significantly higher than the measured switching current, which is affected by thermal escape processes when biased. However, we stress that the value to be used in our modeling is $I_{\mathrm{C}}$, since we are leaving the junction unbiased in the heat transport experiment: what matters here is the plasma frequency, that is, the potential curvature at a minimum, which is defined through $I_{\mathrm{C}}$.

\textit{Linear approximation for heat transfer model.} In the loop configuration, the total phase $\delta$ sits indifferently in one of the local minima of the anharmonic effective Josephson potential, which are all at the same energy since there is no applied voltage or current bias. Due to the environment, $\delta$ experiences fluctuations around the local minimum such that its dynamics may be sensitive to the anharmonicity of the potential. A strong anharmonicity can therefore weaken the approximation of the Josephson device as a LC resonator, which is prerequisite to use the filter picture based on an effective inductor used in the Main text and explained in the subsection below. One can give a simple magnitude estimate of the fluctuations, starting from Langevin equation for $\delta$.

In the framework of the RCSJ model \cite{Stewart1968}, we approximate the CPT as a single JJ with a modified current-phase relation (i.e. effective potential) in a RC environment and biased only by the current noise $i_{\mathrm{N}}$ due to the series resistor $R=R_1+R_2$, with spectral density $S_I(\omega)=(2\hbar\omega/R)\coth(\hbar\omega/2k_{\mathrm{B}}T)$ [see Supplementary Figure \ref{width}]. 
\begin{figure}
	\centering
	\includegraphics[width=10cm]{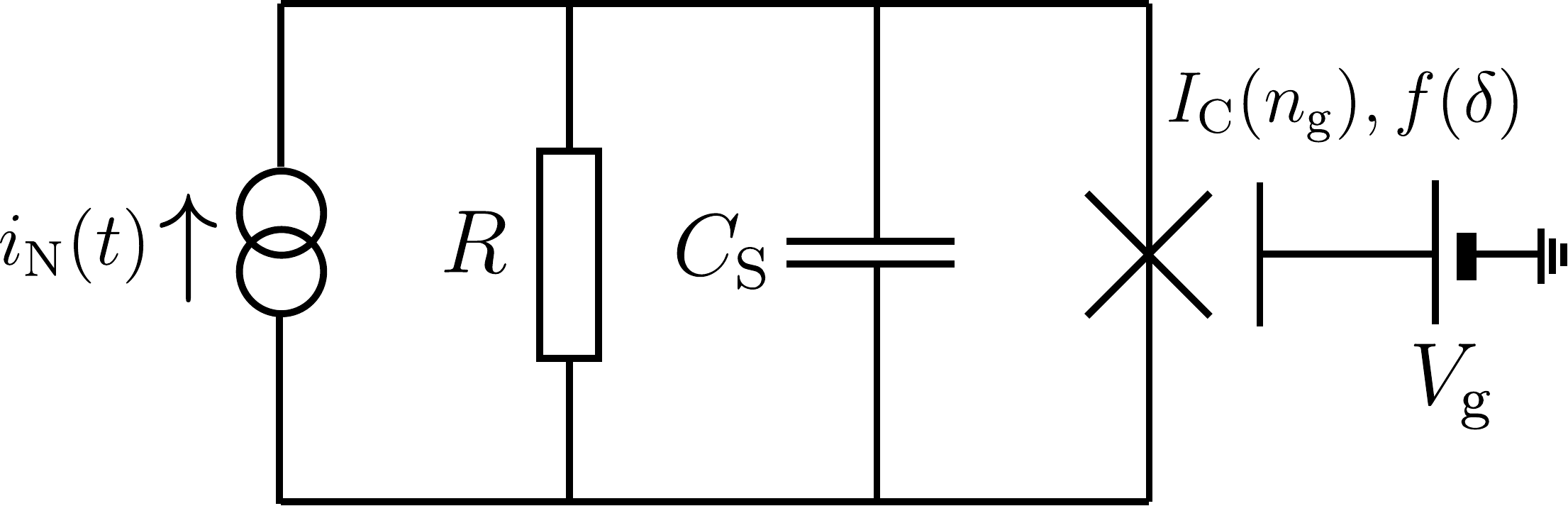}
	\caption{\textbf{RCSJ model representation.} Equivalent circuit in the loop configuration, with $C_{\mathrm{S}}$ the CPT series capacitance, $R$ the series connection of the two resistors generating a Johnson-Nyqvist noise current $i_{\mathrm{N}}(t)$ and the CPT approximated as a single JJ with tunable critical current (/Josephson coupling).}
	\label{width}
\end{figure}

Introducing the inverse time constant $\gamma=1/RC_{\mathrm{S}}$, where $C_{\mathrm{S}}=C/4$ is the series capacitance of the JJ (assuming identical junctions):
\begin{equation}
\label{Langevin}
m\ddot{\delta}+m\gamma\dot{\delta}+f(\delta)=f_{\mathrm{N}}(t),
\end{equation}
where $m=\hbar^2C_{\mathrm{S}}/4e^2$, $f(\delta)=\partial E_0/\partial\delta$, and $f_{\mathrm{N}}(t)=\hbar i_{\mathrm{N}}/2e$. The dynamics of $\delta$ around zero is that of an anharmonic oscillator with effective mass $m$. Linearizing around the equilibrium position $\delta=0$, one obtains the equation of a harmonic oscillator with restoring potential $U(\delta)=m\omega_{\mathrm{p}}^2\delta^2/2$, where $\omega_{\mathrm{p}}=\sqrt{2eI_{\mathrm{C}}/\hbar C_{\mathrm{S}}}$ is the effective JJ plasma frequency. To obtain the typical variance of phase fluctuations, one can make the following observation: in the matched loop configuration, the only capacitance that contributes is that of the series junctions, which is very small ($C_{\mathrm{S}}=$ $0.36$ fF). Therefore, the oscillator is overdamped ($\omega_{\mathrm{p}}/\gamma\sim 0.1$), and $\omega_{\mathrm{p}}\sim 10^{11}$ rad.s$^{-1}$. The resulting hierarchy of typical energies $k_{\mathrm{B}}T\ll\hbar\omega_{\mathrm{p}}\ll\hbar\gamma$ allows us to approximate the noise current in the zero temperature limit ($S_I\approx\hbar\omega/R$) while neglecting the acceleration term $m\ddot{\delta}$ in Supplementary Equation \ref{Langevin}. As a result:
\begin{equation}
\label{variance_int}
\langle\delta^2\rangle \approx\frac{\hbar^3}{2e^2m^2R\gamma^2}\int_0^{\omega_{\mathrm{h}}}\frac{\mathrm{d}\omega}{2\pi}\frac{\omega}{\omega^2+\omega_{\mathrm{p}}^4/\gamma^2}=\frac{R}{R_{\mathrm{Q}}}\ln\left(1+\frac{\omega_{\mathrm{h}}^2}{\omega_{\mathrm{p}}^4/\gamma^2}\right),
\end{equation}
where $R_{\mathrm{Q}}=\pi\hbar/2e^2\approx 6.45~\mathrm{k}\Omega$ is the superconducting resistance quantum and $\omega_{\mathrm{h}}$ is a cut-off frequency that is set to $\omega_{\mathrm{h}}=\gamma$ so as to satisfy the approximation $m\ddot{\delta}\approx 0$. One then obtains a good estimate, for $\gamma/\omega_{\mathrm{p}}\gg 1$, of the typical spreading of the phase fluctuations:
\begin{equation}
\label{equipartition}
\langle\delta^2\rangle\approx\frac{4R}{R_{\mathrm{Q}}}\ln\frac{\gamma}{\omega_{\mathrm{p}}},
\end{equation}
which is small for a low impedance environment $R\ll R_{\mathrm{Q}}$, hence the approximation made in the Main text.

\textit{Landauer formula for the heat flow and thermal conductance.} The spectrum of the Johnson-Nyqvist voltage noise $u_i$ of the resistor $R_i$ at temperature $T_i$ writes:
\begin{equation}
\label{JN noise}
S_{u_i}(\omega)=2R_i\hbar\omega\coth\left(\frac{\hbar\omega}{2k_{\mathrm{B}}T_i}\right),
\end{equation}
where we have included the quantum statistical cut-off that is a consequence of the Bose distribution of thermal photons. In the classical limit one retrieves the original expression $S_{u_i}(\omega)=4k_{\mathrm{B}}T_iR_i$. This voltage noise is modeled as a source in series with the resistor [see Fig. 1c) of the main text]. If the circuit is electrically closed, with a global series impedance $Z_{\mathrm{tot}}(\omega)=R_1+R_2+Z_{\mathrm{ext}}(\omega)$, the noise current spectrum due to resistance $R_i$ flowing into the loop writes:
\begin{equation}
\label{noise_curr}
S_{I_i}(\omega)=\frac{S_{u_i}(\omega)}{|Z_{\mathrm{tot}}(\omega)|^2}.
\end{equation}
The Joule power per unit bandwidth dissipated in resistor $j$ due to the noise current emitted by resistor $i$ is $\dot{q}_{i\rightarrow j}(\omega)=R_jS_{I_i}(\omega)$. One can write equally the heat dissipated in the reverse direction $\dot{q}_{j\rightarrow i}=R_iS_{I_j}(\omega)$. After integrating over all frequencies the resulting net Joule power per unit bandwidth $\dot{q}=\dot{q}_{1\rightarrow 2}-\dot{q}_{2\rightarrow 1}$, one obtains the net average heat power $\dot{Q}_{\gamma}$ due to thermal photons exchanged between the two resistors:
\begin{equation}
\label{heat_landauer}
\dot{Q}_{\gamma}=\int_0^{\infty}\frac{\mathrm{d}\omega}{2\pi}\tau(\omega)\hbar\omega\left[n_1(\omega)-n_2(\omega)\right],
\end{equation}
where $n_i(\omega)=1/\left[\exp(\hbar\omega/k_{\mathrm{B}}T_i)-1\right]$ is the Bose distribution and $\tau(\omega)=4R_1R_2/|Z_{\mathrm{tot}}(\omega)|^2$ is the transmission coefficient of the circuit connecting the two resistors. Using the linear approximation as justified above, the total circuit impedance explicitly writes $Z_{\mathrm{tot}}(\omega,n_{\mathrm{g}})=R_1+R_2+[\mathrm{i}C_{\mathrm{S}}\omega+1/\mathrm{i}L_{\mathrm{J}}(n_{\mathrm{g}})\omega]^{-1}$. We can make a further convenient simplification: the plasma frequency for the whole $n_{\mathrm{g}}$ range is about $10^{11}$ rad.$s^{-1}$, i.e. $\hbar\omega_{\mathrm{p}}\gg k_{\mathrm{B}}T$ throughout all the experiment. As a result, in the angular frequency range of interest (up to $\sim k_{\mathrm{B}}T/\hbar\sim 2\pi\times3$ GHz), the capacitive component is irrelevant, and one may retain the Josephson inductance $L_{\mathrm{J}}$ alone to less than a percent accuracy. The series impedance thus simplifies, $Z_{\mathrm{tot}}(\omega)\approx R_1+R_2+\mathrm{i}L_{\mathrm{J}}\omega$, which leads to the definition of a gate-tunable cut-off frequency for thermal radiation imposed by low-pass filtering, $\omega_{\mathrm{c}}(n_{\mathrm{g}})=(R_1+R_2)/L_{\mathrm{J}}(n_{\mathrm{g}})$. On the other hand, for a small temperature difference between $R_1$ and $R_2$ and introducing the mean temperature $T_{\mathrm{m}}=(T_1+T_2)/2$, one has:
\begin{equation}
\label{approx_bose}
n_1(\omega)-n_2(\omega)\approx\frac{\hbar\omega(T_1-T_2)}{k_{\mathrm{B}}T_{\mathrm{m}}^2}e^{\hbar\omega/k_{\mathrm{B}}T_{\mathrm{m}}}n_{\mathrm{m}}^2(\omega),
\end{equation}
where $n_{\mathrm{m}}$ refers to Bose factor at temperature $T_{\mathrm{m}}$. Introducing $x=\hbar\omega/k_{\mathrm{B}}T_{\mathrm{m}}$, and the reduced circuit cut-off frequency $x_{\mathrm{c}}=\hbar\omega_{\mathrm{c}}/k_{\mathrm{B}}T_{\mathrm{m}}$, one finally obtains the photon heat conductance $G_\gamma\equiv\dot{Q}_{\gamma}/(T_1-T_2)$ presented in the main text:
\begin{equation}
\label{cond_theo_explicit}
G_{\gamma}=\frac{2k_{\mathrm{B}}^2T_{\mathrm{m}}R_1R_2}{\pi\hbar(R_1+R_2)^2}\int_0^{\infty}\mathrm{d}x\frac{x^2e^x}{(e^x-1)^2}\frac{1}{\displaystyle 1+x^2/x_{\mathrm{c}}^2},
\end{equation}
which depends only on universal constants or independently measured parameters.

\section*{Supplementary Discussion}
In our analysis we have neglected quasiparticle excitations, assuming the island was free of any of it and taking the $2e$ periodicity of our thermal conductance oscillations as an evidence that they do not play a crucial role. Below 200 mK and since $\Delta>E_{\mathrm{c}}$ \cite{Matveev1993}, equilibrium quasiparticles are not expected in the island because of the significant even-odd free energy cost of adding an upaired excitation in the island, $F=-k_{\mathrm{B}}T\ln\left[\tanh\left(D(E_{\mathrm{F}})V_{\mathrm{s}}\sqrt{2\pi k_{\mathrm{B}}T\Delta}e^{-\Delta/k_{\mathrm{B}}T}\right)\right]$ \cite{tinkham2004} with $D(E_{\mathrm{F}})=2.15\times 10^{47}$ J$^{-1}$.m$^{-3}$ the density of states at Fermi level for normal Aluminum and $V_{\mathrm{s}}$ the island volume. At a temperature $T^*\approx\Delta/k_{\mathrm{B}}\ln(\Delta D(E_{\mathrm{F}})V_{\mathrm{s}})\approx 250$ mK, $F$ essentially vanishes and thermal excitations completely spoil the $2e$-periodicity. However, even at our lowest working temperatures (or below in many experiments), an excess, non-equilibrium quasiparticle population possibly due to e.g. stray microwave radiation or non-equilibrium phonons with energies $\gtrsim\Delta$ is commonly observed \cite{Ferguson2006,Martinis2009,Mannila2019}, resulting in severe limitations to the coherence of various superconducting devices \cite{Martinis2009,Catelani2011,Riste2013}. We use the three-level model of Supplementary Reference \cite{Aumentado2004}, where the states corresponding to zero quasiparticle near or in the island (even parity), to one quasiparticle in the leads near the island (even parity) and to one quasiparticle in the island (odd parity) are, respectively: 
\begin{multline}
\\
\epsilon_0=E_0(n_{\mathrm{g}},\delta),\\
\epsilon_l=E_0(n_{\mathrm{g}},\delta)+\Delta,\\
\epsilon_i=E_0(n_{\mathrm{g}}+1,\delta)+\Delta,\\
\end{multline}
We assume that no Cooper pair is broken in the island due to its small volume. In addition, the two Al layers were evaporated with nominally identical settings, so we assume the gaps in the island and in the leads to be equal. Furthermore, we assume a temperature-independent pair breaking rate $\Gamma_{0l}$ in the regime where non-equilibrium QP dominate (typically well below 250 mK). We call $\delta\epsilon(n_{\mathrm{g}},\delta)=\epsilon_l-\epsilon_i=E_0(n_{\mathrm{g}},\delta)-E_0(n_{\mathrm{g}}+1,\delta)$ the energy difference between the even and the odd parity in presence of a non-equilibrium quasiparticle. In the range $n_{\mathrm{g}}$ mod 2 $\in[-0.5,0.5]$, $\delta\epsilon>0$, hence the island is seen as an energy barrier for QP, that cannot be easily overcome at temperatures $T\ll\delta\epsilon/k_{\mathrm{B}}$. As a result, the probability for a parity switch to occur in this interval is vanishing at low temperatures. On the contrary, in the range $n_{\mathrm{g}}$ mod 2 $\in[0.5,1.5]$, $\delta\epsilon<0$. The island is therefore energetically a "trap" for a quasiparticle excitation: after a pair breaking event creating a quasiparticle near the CPT, a tunneling of the QP in the island will be favourable energetically. As a result, an unpaired excitation may tunnel into the island at a rate $\Gamma_{li}$ (even to odd parity switch) faster than our measurement times. Such an event leads to a shift by an amount $e$ of the gate charge seen by the noise current flowing through the CPT, after which the quasiparticle may tunnel out by thermal activation at a rate $\Gamma_{il}=\Gamma_{li}e^{\delta\epsilon/k_{\mathrm{B}}T}$. At temperatures $T\ll\delta\epsilon/k_{\mathrm{B}}$, a QP that has entered the island could thus be trapped there for a long time and substantially affect the parity occupation probabilities.
\begin{figure}
	\centering
	\includegraphics[width=8cm]{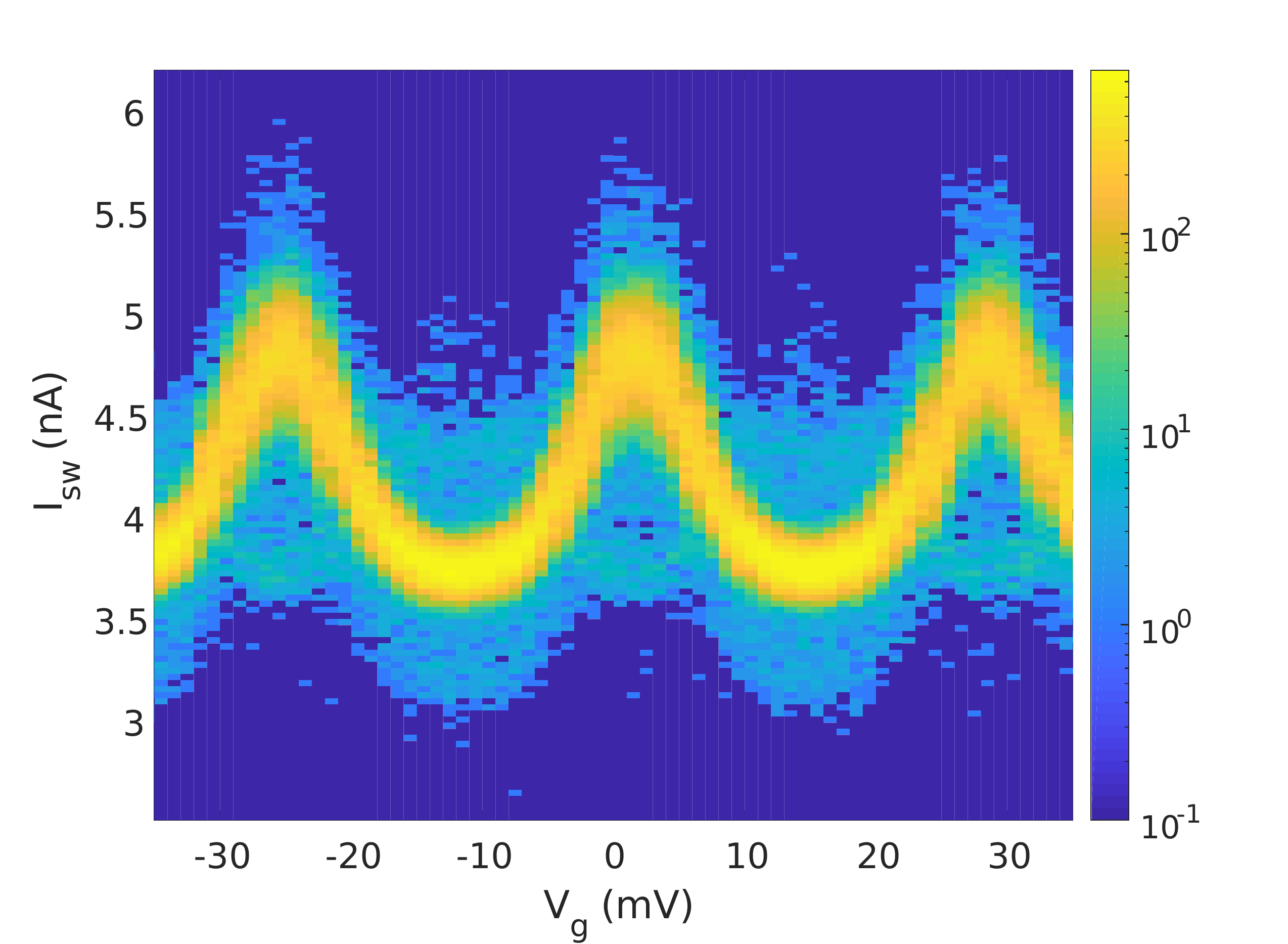}
	\includegraphics[width=8cm]{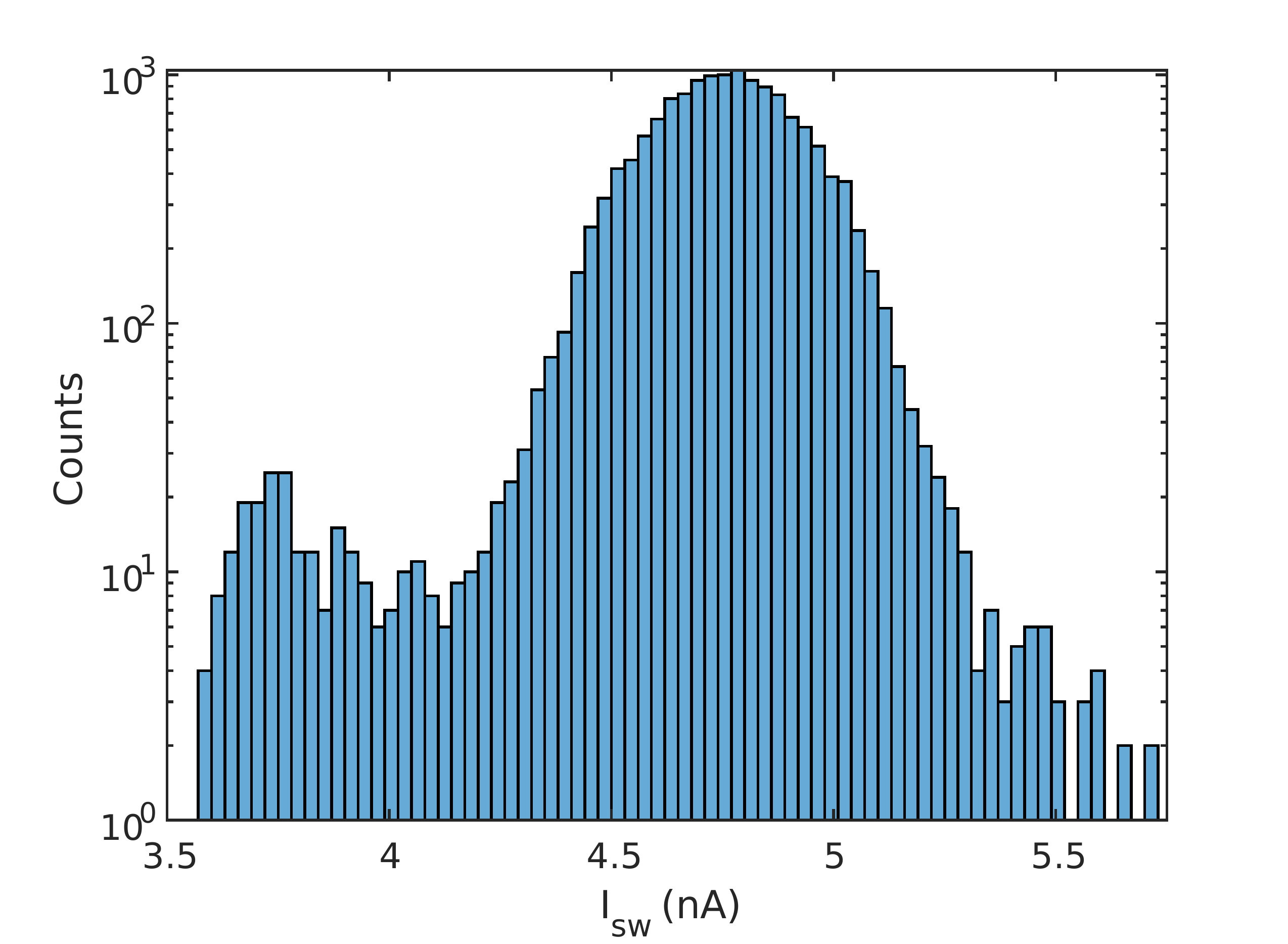}
	\caption{\textbf{Switching current histograms.} a) Colormap of switching current histograms versus applied gate voltage at 100 mK, for a ramp rate of $10~\mu$A.s$^{-1}$. The logarithmic scale is intended to reveal poisoning events (the "ghost trace" below each peak), that are barely visible with a linear scale. Each cut is a histogram of switching currents with 5000 attempts. b) Histogram of switching events for $n_{\mathrm{g}}=1$ mod 2, with data collected from the three cuts at peak maxima.}
	\label{colormap_Isw}
\end{figure}
To assess qualitatively the magnitude of quasiparticle poisoning in our device, we have measured in the characterization run (no loop) the switching current of the CPT as a function of the gate voltage for several temperatures \cite{Joyez1994,Mannik2004,Aumentado2004}. To obtain a good estimate of the even/odd occupation probabilities, we ramp the current flowing through the CPT beyond the switching current with rates ranging from $100$ nA.s$^{-1}$ to $10~\mu$A.s$^{-1}$. If the ramp is fast enough, then by repeating the procedure $\sim 10^3-10^4$ times, we obtain a "snapshot" of the even and odd states occupation, and the switching current distribution typically features a bi-modal distribution. In Supplementary Figure \ref{colormap_Isw}b) we have an example of such a distribution, yet it is difficult to obtain with a good accuracy the odd state occupancy due to the significantly low number of corresponding switching events. Besides, the switching current technique suffers from its limited bandwidth set by the finite time of switching to the voltage state of the CPT (here measured to be around 500 $\mu$s for all temperatures), which results in several events that lie in between $I_{\mathrm{sw,odd}}$ and $I_{\mathrm{sw,even}}$. In addition, our elevated operation temperatures cause broadening of the histograms \cite{barone1982physics}, which limits further our ability to put a clear separation between odd and even parity events. Therefore, a quantitative comparison with the model developed in Supplementary Reference \cite{Aumentado2004} is not possible for our results. Nevertheless, we can put a lower bound of 96 $\%$ on the even number occupation probability below 200mK.

The observed bimodal behaviour of the switching currents distribution at a ramp rate of $10~\mu$A.s$^{-1}$ is washed out for lower ramp rates: in such a situation, a poisoning event might occur in between the switching current expected for the odd parity and the one expected for the even parity. This allows us to extract a characteristic poisoning time $\Gamma_{li}^{-1}$ in the $100~\mu$s range for odd integer $n_{\mathrm{g}}$, where it is expected to be the shortest due to the "QP trap" configuration of the CPT. Since the energy difference $\delta\epsilon$ between parities is typically $E_{\mathrm{J}}/2\sim 2-3.5k_{\mathrm{B}}T$ at odd integer $n_{\mathrm{g}}$ for our temperatures, such a trap is rather shallow and we can estimate that a QP dwells in the island over a typical time $\Gamma_{il}^{-1}$ in the ms range before tunneling out.

For temperatures approaching 200 mK and above, the free energy cost of adding one quasiparticle in the island is reduced enough for a significant occupation probability of the odd parity state, and poisoning events occur at timescales faster than the switching time. As a result, the switching currents will be weighted auto-averages of switching currents in even and odd states, which here manifests through additional peaks at even integer $n_{\mathrm{g}}$ values. Full $1e$ periodicity of the switching current is recovered at $T=250$ mK (see Supplementary Figure \ref{poison}), which corresponds to the temperature $T^*$ where the even-odd free energy cost vanishes.

\begin{figure}
	\centering
	\includegraphics[width=10cm]{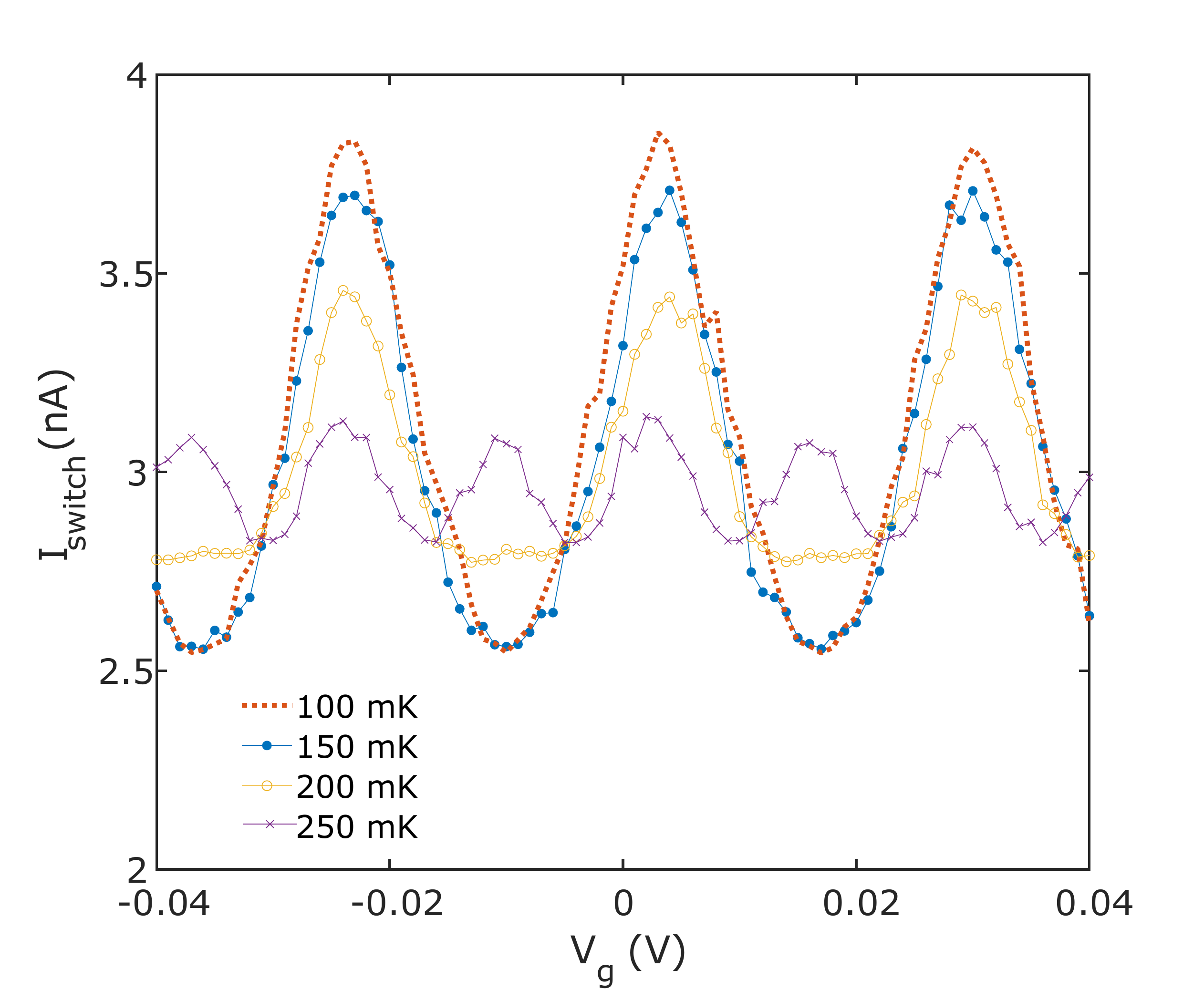}
	\caption{\textbf{Switching current modulation versus applied gate voltage for different cryostat temperatures.} Each point is the average of 100 switching events.}
	\label{poison}
\end{figure}

We now turn to consequences of poisoning on our heat transport measurements. The cut-off frequency $\omega_{\mathrm{c}}=(R_1+R_2)/L_{\mathrm{J}}$ for blackbody radiation lies in the GHz range for all gate positions, and is therefore several orders of magnitude larger than the typical non-equilibrium poisoning rates $\Gamma_{li},\Gamma_{il}$ based on our estimates. As a result, the low-pass filtering characteristic is unchanged by parity switches for noise current frequencies comparable with or smaller than the typical poisoning rates. At frequencies of the emitted radiation in the decade below $\omega_{\mathrm{c}}$, which weight the most in the radiation spectrum, the noise current flows through the CPT much faster than the parity switches. As a result, it should "sample" well both parity states, and the measured thermal conductance should be a weighted average of its value at $n_{\mathrm{g}}$ and that at $n_{\mathrm{g}}+1$, with weights corresponding to even and odd state probabilities, respectively. Based on our measurements (96 to 98 $\%$ probability of having even parity at $n_{\mathrm{g}}=1$) and the fact that the poisoning is more likely for odd integer $n_{\mathrm{g}}$ (therefore making those numbers a low bound for even parity probability regardless of $n_{\mathrm{g}}$), we conclude that poisoning by nonequilibrium QP plays no significant role in our photonic heat transport measurements. Meanwhile, the onset of equilibrium poisoning around 200 mK may explain in part the reduction in the contrast of thermal conductance modulation when $T_{\mathrm{m}}$ approaches 200 mK.

\end{document}